\documentclass[aps,twocolumn,showpacs,preprintnumbers,noshowpacs,superscriptaddress,6pt]{revtex4-2}
\pdfoutput=1
\pdfoutput=1
\pdfoutput=1
\pdfoutput=1
\pdfoutput=1
\usepackage[utf8]{inputenc}
\usepackage{graphicx,epstopdf}
\usepackage{subfigure}
\usepackage{multirow}
\usepackage{adjustbox}
\usepackage{natbib}

\usepackage{longtable}
\usepackage{parskip}
\usepackage[T1]{fontenc}
\usepackage{dcolumn}

\usepackage{bm}
\usepackage{amsfonts}
\usepackage{amsmath}
\usepackage{amssymb}
\usepackage{siunitx}
\usepackage{booktabs}
\usepackage{physics}
\usepackage{mathtools}
\usepackage{flafter}
\usepackage{threeparttable}
\usepackage{subfiles}
\usepackage{subfigure}
\usepackage{import}
\usepackage{float}
\usepackage{makecell}
\usepackage{diagbox}

\usepackage{ulem}

\usepackage{color, colortbl}
\definecolor{Gray}{gray}{0.9}
\definecolor{LightCyan}{rgb}{0.88,1,1}
\usepackage{indentfirst}
\usepackage{mdframed}
\usepackage[strict]{chngpage}

\newmdenv{allfour}

\newmdenv[leftline=false,rightline=false]{topbot}

\newmdenv[topline=false,rightline=false]{leftbot}

\usepackage{color}
\definecolor{BLUE}{rgb}{0.0,0.0,1.0}

\begin{document}

\title{Ground state of superheavy elements with $120 \leq Z \leq 170$: systematic study of the electron-correlation, Breit, and QED effects}

\author{I.~M.~Savelyev}
\affiliation{Department of Physics, St. Petersburg State University, 7/9 Universitetskaya nab., 199034 St. Petersburg, Russia}

\author{M.~Y.~Kaygorodov}
\affiliation{Department of Physics, St. Petersburg State University, 7/9 Universitetskaya nab., 199034 St. Petersburg, Russia}

\author{Y.~S.~Kozhedub}
\affiliation{Department of Physics, St. Petersburg State University, 7/9 Universitetskaya nab., 199034 St. Petersburg, Russia}

\author{A.~V.~Malyshev}
\affiliation{Department of Physics, St. Petersburg State University, 7/9 Universitetskaya nab., 199034 St. Petersburg, Russia}

\author{I.~I.~Tupitsyn}
\affiliation{Department of Physics, St. Petersburg State University, 7/9 Universitetskaya nab., 199034 St. Petersburg, Russia}

\author{V.~M.~Shabaev}
\affiliation{Department of Physics, St. Petersburg State University, 7/9 Universitetskaya nab., 199034 St. Petersburg, Russia}
\affiliation{National Research Centre “Kurchatov Institute” B.P. Konstantinov Petersburg Nuclear Physics Institute, Gatchina,
Leningrad district 188300, Russia}

\date{\today}

\begin{abstract}
For superheavy elements with atomic numbers $120\leq Z \leq 170$, the concept of the ground-state configuration is being reexamined. 
To this end, relativistic calculations of the electronic structure of the low-lying levels are carried out by means of the Dirac-Fock and configuration-interaction methods.
The magnetic and retardation parts of the Breit interaction as well as the QED effects are taken into account. 
The influence of the relativistic, QED, and electron-electron correlation effects on the determination of the ground-state is analyzed.
\end{abstract}

\maketitle
\section{Introduction}\label{sec:intro} 
Mendeleev's Periodic Table is an empirically supported scheme which allows one to categorize the physical and chemical properties   of the elements by linking them with the rule of the successive occupation of the atomic orbitals. 
With increasing the atomic number~$Z$, relativistic effects grow substantially.
They can significantly alter various properties of the elements as compared with their lighter homologues.
A classical example of the relativistic effects is the yellow color of gold~\cite{1979_Pyykko_ACR, 1988_PyykkoP_CR88, 2012_PyykkoP_CR112}.
In the region of the superheavy elements (SHEs) belonging to the $7$th period of the Table, the interplay between the relativistic and electron-electron correlation effects gives rise to the trends of deviation from the periodic law~\cite{2015_OganessianY_RepProgPhys, 2017_OganessianY_PhysScrip, 2015_PERSHINA, 2019_Pershina, 2019_GiulianiS_RevModPhys}.
Some of these deviations, such as a different ground-state configuration of Lr~$(Z=103)$ relative to its lighter homologue Lu~$(Z=71)$, are confirmed experimentally~\cite{2015_SatoT_Nature}, the others, like a positive electron affinity in Og~$(Z=118)$, are predicted only theoretically~\cite{1996_EliavE_PhysRevLett, 2003_GoidenkoI_PhysRevA, 2018_LackenbyB_PhysRevA98Og, 2021_GuoY_AQC83, 2021_KaygorodovM_PhysRevA}.
Whether the~$8$th-period SHEs (with~$Z>118$) fit the Periodic Table and obey the periodic law is an open  intriguing question.
For instance, this period brings into play the previously-never-met 5g shell, and the corresponding electronic-structure feature no doubt must be presented in possible extensions of the Periodic Table.
In addition, the influence of the quantum-electrodynamics (QED) effects on the SHE ground states has not been investigated systematically so far.
\par
A review of the current status of the problem and an extension of the Periodic Table up to $Z = 172$ based on the Dirac-Fock (DF) calculations, also known as the relativistic Hartree-Fock ones, are presented in Ref.~\cite{2010_PykkoP_PhysChemChemPhys13}, see also a very recent review~\cite{Smits:HAL:2022}.
This upper bound is determined by the fact that at higher values of Z the lowest (1s) Dirac level ``dives'' into the negative-energy continuum, provided a reasonable model for the nuclear charge distribution is employed~\cite{1945_PomaranchukI_JPUSSR9, 1961_Voronkov_ru, 1969_Gershtein_ru, *1969_Gershtein_nuova, 1969_PieperW_ZPHN218,  Zeldovich:1971ru, 1985_GreinerW_QEDStrongFields, Popov:2001ru, 2017_GodunovS_EPJC, 2017_Rafelski_springer, 2020_PopovR_PRD, 2021_Voskresensky, 2022_Sveshnikov_MPLA}. 
\par
The key point for the description of the SHE properties is determination of the ground-state configuration.
The first attempts to advance the study of the SHEs beyond the $7$th period and to conjecture on their chemical and physical properties were undertaken in the $1970$s~\cite{1970_MannJ_JChemPhys, 1971_LuC_ADNDT3, 1971_FrickeB_ThChimAct21, 1972_FrickeB_JCP57, 1973_Desclaux_ADNT12, 1977_FrickeB_ADNDT19}.
Throughout the years, the issue was addressed by using various one-configuration methods~\cite{1996_UmomotoK_JPSJ, 2017_ZhouZ_ADNDT114}.
It soon became clear that in many cases the total energies of various configurations differ very little from each other, and more sophisticated configuration-interaction calculations are necessary. 
Taking into account the correlation effects may lead to a change in the ground-state configuration. 
\par
Some SHEs from the $8$th period were studied within the many-configuration approaches~\cite{1998_EliavE_JCP109, 2002_EliavE_JPhysB, 2005_Lim, 2007_IndelicatoP_EurPhysJD,  2011_IndelicatoP_TCA129, 2013_Scripnikov, 2013_Borschevsky, 2016_GingesJ_PhysRevA, *2016_GingesJ_JPhysBAtMolOptPhys, 2016_Zaitsevskii, 2021_TupitsynI_OS129_ru}.
The only paper that went beyond the one-configuration approximation for all the $8$th period elements is Ref.~\cite{2006_NefedovV_DoklPhysChem}. The multiconfiguration Dirac–Fock method was used there to account for the interaction between energetically close configurations in the SHEs with~$Z\leq 164$.
As a result, in about~$50\%$ of cases the ground-state configurations found in Ref.~\cite{2006_NefedovV_DoklPhysChem}  differ from the ones obtained in the previous Dirac-Fock-Slater calculations~\cite{1977_FrickeB_ADNDT19}, where no electron-electron correlation effects were considered.
In Ref.~\cite{2006_NefedovV_DoklPhysChem}, only the ground-state configurations were reported without any information on the level structure. 
The stability of the obtained results with respect to the accuracy of the correlation treatment was not discussed in that work as well.
\par
There are also some other issues that need to be clarified when discussing the SHE ground states.
Does the one-configuration description remain valid for so complex systems possessing quite a large number of valence shells with, in particular, the $5g$ one among them?
In other words, it seems reasonable that the ground-state level of the SHEs generally can not be found without taking into account the electron-correlation effects, but is it possible, in principle, to describe with a sufficient accuracy this state using a single configuration?
Can previously unaccounted QED effects change the ground state of the SHEs?
The present paper aims to investigate these points in the framework of the relativistic Dirac-Fock method and the configuration-interaction method in the basis of the Dirac-Fock-Sturm orbitals~\cite{2003_TupitsynI_PhysRevA, 2005_TupitsynI_PhysRevA, 2018_TupitsynI_PhysRevA}.
In our calculations, in order to investigate a possible influence of the QED effects on the electronic structure and the ground-state configuration, both the methods are paired with the model-QED-operator approach~\cite{2013_ShabaevV_PhysRevA, 2015_ShabaevV_CompPhysComm, *2018_ShabaevV_CompPhysComm} which has been recently extended to the region $120\leq Z \leq 170$ in Ref.~\cite{2022_MalyshevA_PRA106}.
\par
The paper is organized as follows.
In Sec.~\ref{sec:theory}, an overview of the methods and their main implementation features are presented.
The numerical details and the particular aspects of the calculation procedure are given in Sec.~\ref{sec:details}.
We discuss the obtained results and compare them with the previous calculations in Sec.~\ref{sec:results}.
Sec.~\ref{sec:conclusion} concludes the paper with a brief summary.
\par
The atomic units are used throughout the paper.

\section{Theoretical approaches and methods}\label{sec:theory}
In the present work, to explore the SHE ground states we use the Dirac-Fock (DF) and configuration-interaction (CI) methods. The issue is studied from the different perspectives using the one- as well as the many-configuration approaches.

We consider a relativistic configuration $K$ defined by the occupation numbers $\{q_a\}_{a=1}^{N_s}$, where the index $a$ enumerates the relativistic shells. For the list of the relativistic shells $(n_1 l_1 j _1)^{q_1} \ldots (n_{N_s} l_{N_s} j_{N_s})^{q_{N_s}} $ ($n$ is the principal quantum number, $l$ and $j$ are the orbital and total angular momenta, respectively), we first determine the DF energy obtained within the relativistic-configuration-average (RAV) approximation, also known in the literature as the $jj$-average one \cite{1975_Lindgren}. The corresponding total DF energy can be formally written as
\begin{align}
E_{\rm RAV}^{\rm DF}(K) = 
\frac{1}{N_d} \sum_{\alpha\in K} 
\langle \alpha | \hat{H}^{\rm DC} | \alpha \rangle \, ,
\end{align}
where $\alpha \equiv {\rm det}_\alpha\{\varphi_i^{\rm DF}\}$ are the Slater determinant constructed from the one-electron DF orbitals~$\varphi_i^{\rm DF}$ belonging to the configuration~$K$, $N_d$ is the number of these determinants, and $\hat{H}^{\rm DC}\equiv \hat{H}^{\rm D} + \hat{V}^{\rm C}$ is the many-electron Dirac-Coulomb Hamiltonian. In the DF-RAV approximation, the summation in the functional~(1), which can be performed analytically, is equivalent to the summation over the relativistic terms~$J$ of the configuration~$K$ taking into account their multiplicities~\cite{1961_GrantI_PRSL262,  1970_GrantI_AP19}. The DF equations can be derived by varying Eq.~(1) with the evident constraints due to the orthonormality conditions. The DF orbitals~$\varphi_i^{\rm DF}$ and the energy~$E_{\rm RAV}^{\rm DF}(K)$ are then obtained by solving the corresponding DF equations. We note that only the Coulomb-interaction operator~$\hat{V}_{\rm C}$ is included self-consistently into the DF equations.

The Breit-interaction correction to the average energy~$E_{\rm RAV}^{\rm DF}(K)$ of the configuration $K$ is evaluated perturbatively as
\begin{align}\label{eq:de_breit}
\Delta E_{\rm RAV}^{\rm B}(K) = 
\frac{1}{N_d} \sum_{\alpha\in K} 
\langle \alpha | \hat{V}^{\rm B} | \alpha \rangle \, ,
\end{align}
where the determinants~$\alpha$ are constructed from~$\varphi_i^{\rm DF}$ and $\hat{V}^{\rm B}$ is the Breit-interaction operator. The QED corrections are treated using the model-QED-operator approach developed in Refs.~\cite{2013_ShabaevV_PhysRevA, 2015_ShabaevV_CompPhysComm, *2018_ShabaevV_CompPhysComm, 2022_MalyshevA_PRA106}. We mention also some alternative methods to approximately account for the QED effects in many-electron systems, see, e.g., Refs.~\cite{1990_Indelicato, 2003_Pyykko_JPB36, 2003_Draganic, 2005_FlambaumV_PhysRevA, 2007_IndelicatoP_EurPhysJD, 2010_ThierfelderC_PhysRevA, 2013_TupitsynI_OptSpectrosc_ru, 2016_GingesJ_PhysRevA, *2016_GingesJ_JPhysBAtMolOptPhys}. 
For very recent applications and developments of the model-QED-operator methods, which include the calculations of the QED effects in molecules, we refer to Refs.~\cite{2021_Skripnikov_JCP, 2021_Skripnikov_Chubukov_JCP, 2021_Sunaga, 2022_Sunaga, 2022_Oleynichenko}.
Like the Breit-interaction contribution, the QED correction in the RAV approximation is calculated as the relativistic-configuration-average expectation value of the model-QED operator $\hat{V}^{\rm Q}$,
\begin{align}\label{eq:de_qed}
\Delta E_{\rm RAV}^{\rm Q}(K) = 
\frac{1}{N_d} \sum_{\alpha\in K} 
\langle \alpha | \hat{V}^{\rm Q} | \alpha \rangle \, .
\end{align}
Finally, for the configuration $K$, we consider the average energy including the Breit correction and QED effects,
\begin{align}
E_{\rm RAV}^{\rm DCBQ}(K) = 
E_{\rm RAV}^{\rm DF}(K) + \Delta E_{\rm RAV}^{\rm B}(K) + \Delta E_{\rm RAV}^{\rm Q}(K) \, .
\end{align}
Hereinafter, this scheme is referred to as the DCBQ-RAV one. 

To resolve the level structure for the configuration~$K$, one can try to find a single-configuration DF wave function and corresponding energy for the $jj$-coupling term using the energy functional constructed for a given value of~$J$. However, if this approach is employed for an open-shell system possessing a complex level structure, it often proves impossible to adequately select a proper linear combination of many-electron wave functions solely from symmetry considerations. It turns out that most of the SHEs have several open shells, and, therefore, this straightforward scheme may result in a wrong level structure. In the present work, the level structure of the configuration~$K$ is resolved by means of the CI approach for the Dirac-Coulomb-Breit~(DCB) Hamiltonian supplemented with the model-QED operator. The DCBQ Hamiltonian reads as
\begin{align}\label{eq:H^DCBQ}
\hat{H}^{\rm DCBQ} = 
\Lambda^{+}
\left[
\hat{H}^{\rm D} + \hat{V}^{\rm C} + \hat{V}^{\rm B} + \hat{V}^{\rm Q}
\right] 
\Lambda^{+} \, ,
\end{align}
where $\Lambda^{+}$ is the product of one-electron projectors on the positive-energy solutions of the DF-RAV equations. The eigenvalue problem induced by the Hamiltonian~(5) in the space of all the determinants~$\alpha$ arising from the single relativistic configuration (SRC)~$K$ describes the splitting of the levels,
\begin{align}\label{eq:e_dcbq_src}
\hat{H}^{\rm DCBQ} 
\Psi_{\rm SRC}(K,JM) = 
E^{\rm DCBQ}_{\rm SRC} 
\Psi_{\rm SRC}(K,JM) \, ,
\end{align}
where $M$ means the projection of $J$.

However, in case of energetically close configurations, their strong interaction and mixing may result in changes of the level structure. To account for the correlation effects, we consider a larger CI problem,
\begin{align}\label{eq:dcbq_ci}
\hat{H}^{\rm DCBQ} 
\Psi_{\rm CI}(JM) =
E^{\rm DCBQ}_{\rm CI}(J)
\Psi_{\rm CI}(JM) \,,
\end{align}
in the space spanned by the Slater determinants generated not only from the configuration~$K$ but also from a given list of relativistic configurations (see details in the next section). In the present calculations, the CI method in the basis of the Dirac-Fock-Sturm orbitals is used (CI-DFS)~\cite{2003_TupitsynI_PhysRevA,2005_TupitsynI_PhysRevA,2018_TupitsynI_PhysRevA}. 
At the CI level, the Breit and QED corrections are taken into account, according to Eqs.~(\ref{eq:H^DCBQ}) -- (\ref{eq:dcbq_ci}), by including the corresponding terms into the Dirac-Coulomb Hamiltonian.

Finally, we emphasize that the main goal of the present study is not to obtain the most accurate theoretical predictions for the SHE energy-level structure, since in the cases of complex configurations this can be a separate extremely complicated task. Instead, we aim at a reliable determination of the ground-state levels and the configurations they belong to within a series of the adequate relativistic calculations.

Having discussed the methods, let us proceed to details of their application in the scope of the present work.

\section{Details of the calculations}\label{sec:details}
In the present work, all the calculations of the energy levels of SHE are performed employing the Fermi model for the nuclear-charge distribution. The root-mean-square radius of the nucleus (in fm) is given by 
\begin{equation}
    R = \sqrt{\frac{3}{5}}R_{\mathrm{sphere}}, \qquad R_{\mathrm{sphere}} = 1.2 A^{1/3},
\end{equation}
where for the nucleon number $A$ we use the approximate formula from Ref.~\cite{1969_PieperW_ZPHN218}, \begin{equation}\label{eq:nucleon}
    A = 0.00733 Z^2 + 1.30 Z + 63.6.
\end{equation} The value of $A$ obtained from Eq.~(\ref{eq:nucleon}) is rounded to the nearest integer. This choice of the nuclear size is consistent with the one made in Ref.~\cite{2022_MalyshevA_PRA106}.

The SHE ground-state configuration is \textit{a~priori} unknown. 
As  described in the previous section, we use three schemes to define the ground-state configuration. 
The first scheme is based on the DF-RAV method.
Probing various configurations $K$, we determine the ground-state one as the configuration~$K^{\ast}$ with the lowest average energy~$E^{\mathrm{DCBQ}}_{\mathrm{RAV}}(K^{\ast})$.
For each~$Z$, the list of all possible configuration-candidates for the role of the ground one is constructed by distributing~$N_e=Z-N_{\mathrm{core}}$ valence electrons over the valence relativistic shells.
The number of the core-shell electrons~$N_{\mathrm{core}}$ and the list of the valence shells are presented in Table~\ref{tab:1}.
We consider as the valence shells those ones which, according to our preliminary calculations, are most likely to be occupied in the ground state.
\begin{table}[H]
\centering
\caption{The list of the relativistic shells used to generate the relativistic configurations for which the DF-RAV equations are solved. The absence of the lower index $j$ in the column ``Core shells''  means that all relativistic orbitals corresponding to the nonrelativistic one are fully occupied.
The probe configurations are generated according to the following rule: the core shells are fully occupied, the $Z-N_{\mathrm{core}}$ valence electrons are distributed over the valence shells.
The notations~$[\mathrm{Rn}]$ and~$[\mathrm{Og}]$ represent the closed-shell configurations of radon and oganesson atoms, respectively.
}
\resizebox{\columnwidth}{!}{%
\begin{tabular}{
c@{\quad}
l@{\quad}
c@{\quad}
l
}

\toprule
\multicolumn{1}{c}{$Z$} & \multicolumn{1}{c}{Core shells} &
\multicolumn{1}{c}{$N_\mathrm{core}$} &\multicolumn{1}{c}{Valence shells}
\\
\midrule
120\,\,--\,121 & $[\mathrm{Rn}] 5f \, 6d \, 7s \, 7p_{1/2}$           & 114 & $7p_{3/2} \, 8s \, 8p_{1/2} \, 7d_{3/2}$  \\
122\,\,--\,123 & $[\mathrm{Og}]$                                      & 118 & $8s \, 8p_{1/2} \, 7d_{3/2} \, 6f_{5/2}$  \\
124\,\,--\,133 & $[\mathrm{Og}] 8s$                                   & 120 & $8p_{1/2} \, 6f_{5/2} \, 7d_{3/2} \, 5g_{7/2}$  \\
134\,\,--\,144 & $[\mathrm{Og}] 8s \, 5g_{7/2}$                       & 128 & $8p_{1/2} \, 6f_{5/2} \, 7d_{3/2} \, 5g_{9/2}$  \\
145\,\,--\,146 & $[\mathrm{Og}] 8s \, 8p_{1/2} \,5g_{7/2}$            & 130 & $6f_{5/2} \, 7d_{3/2} \, 5g_{9/2} \, 9s$ \\
147\,\,--\,155 & $[\mathrm{Og}] 8s \, 8p_{1/2} \,5g$                  & 140 & $6f_{5/2} \, 7d_{3/2} \, 6f_{7/2} \, 9s$ \\
156\,\,--\,160 & $[\mathrm{Og}] 8s \, 8p_{1/2} \,5g \,6f_{5/2}$       & 146 & $6f_{7/2} \, 7d_{3/2} \, 9s \, 7d_{5/2}$ \\
161\,\,--\,165 & $[\mathrm{Og}] 8s \, 8p_{1/2} \,5g \,6f$             & 154 & $7d_{3/2} \, 7d_{5/2} \, 9s \, 8p_{3/2}$ \\
166\,\,--\,168 & $[\mathrm{Og}] 8s \, 8p_{1/2} \,5g \,6f \,7d_{3/2}$  & 158 & $7d_{5/2} \, 9s \,  8p_{3/2} \, 9p_{1/2}$ \\
169\,\,--\,170 & $[\mathrm{Og}] 8s \, 8p_{1/2} \,5g \,6f \,7d$        & 164 & $9s \, 8p_{3/2} \, 9p_{1/2} \, 7f_{5/2}$ \\

\bottomrule
\end{tabular}%
} 
\label{tab:1}
\end{table}

\par
As an example, in Table~\ref{tab:2} for the SHEs with~$Z=125$ and~$Z=140$, seven configurations~$K$ with the lowest relativistic-configuration-average energies~$E^{\mathrm{DF}}_{\mathrm{RAV}}(K)$ are presented.
For each configuration~$K$ given relative to the closed-shell one, the total DF-RAV energy and the energies obtained by successively adding the Breit and QED corrections are placed in the fourth, fifth, and sixth columns of Table~\ref{tab:2}, respectively.
The configurations are sorted in ascending order of the energy~$E^{\mathrm{DF}}_{\mathrm{RAV}}(K)$, i.e., the first entry corresponds to the configuration with the lowest energy.
The relative changes in the order of the configurations after addition of the corrections are indicated by the symbols~$\bigtriangledown$ (down) and $\triangle$ (up). The absence of these symbols corresponds to the case when the configuration position in the sorted list does not change.
\onecolumngrid\
\begin{table}[H]
\centering
\caption{The relativistic-configuration-average energies for the SHEs with~$Z = 125$ and~$Z = 140$ evaluated for the configurations~$K$ using the DF method,~$E^{\mathrm{DF}}_{\mathrm{RAV}}$, with the addition of the Breit-interaction correction,~$+\Delta E^{\mathrm{B}}_{\mathrm{RAV}}$, and with the additional accounting for the QED correction,~$+\Delta E^{\mathrm{Q}}_{\mathrm{RAV}}$, (a.u.).
The configurations are presented relative to the closed-shell configuration and sorted in the ascending order of the energy~$E^{\mathrm{DF}}_{\mathrm{RAV}}$.
In the last two columns, the symbols~$\bigtriangledown$ (down) and~$\triangle$ (up) indicate the change in the order of the configurations relative to the order in the preceding column. 
The absence of these symbols means that there is no change in the position of the configuration. In particular, the QED effects do not affect the order.
}
\renewcommand{\tabcolsep}{0.3cm}
\begin{tabular}{
c
l
l
S[table-format=-5.4]
S[table-format=-5.4]
S[table-format=-8.4]
}

\toprule
\multicolumn{1}{c}{$Z$} &
\multicolumn{1}{c}{Closed Shells}  &
\multicolumn{1}{c}{$K$} &
\multicolumn{1}{c}{$E^{\mathrm{DF}}_{\mathrm{RAV}}$} &
\multicolumn{1}{c}{$+\Delta E^{\mathrm{B}}_{\mathrm{RAV}}$} &
\multicolumn{1}{c}{$+\Delta E^{\mathrm{Q}}_{\mathrm{RAV}}$} \\
\midrule
\multirow{7}{*}{125}  &  \multirow{7}{*}{$\mathrm{[Og]} 8s_{1/2}^2$}  & $8p_{1/2}^1 6f_{5/2}^4                       $ & -64846.3116 & -64718.7639 $\, \bigtriangledown 3$  & -64627.5323 \\
                      &                                      & $8p_{1/2}^1 6f_{5/2}^3 5g_{7/2}^1            $ & -64846.3061 & -64718.7781 $\, \triangle        1$  & -64627.5496 \\
                      &                                      & $8p_{1/2}^1 7d_{3/2}^1 6f_{5/2}^2 5g_{7/2}^1 $ & -64846.3033 & -64718.7718 $\, \triangle        1$  & -64627.5421 \\
                      &                                      & $8p_{1/2}^2 6f_{5/2}^2 5g_{7/2}^1            $ & -64846.3007 & -64718.7680 $\, \triangle        1$  & -64627.5377 \\
                      &                                      & $7d_{3/2}^1 6f_{5/2}^4                       $ & -64846.2769 & -64718.7300 $\, \bigtriangledown 1$  & -64627.4988 \\
                      &                                      & $8p_{1/2}^1 7d_{3/2}^1 6f_{5/2}^3            $ & -64846.2697 & -64718.7182 $\, \bigtriangledown 1$  & -64627.4854 \\
                      &                                      & $7d_{3/2}^1 6f_{5/2}^3 5g_{7/2}^1            $ & -64846.2680 & -64718.7408 $\, \triangle        2$  & -64627.5127  \\
\midrule
\multirow{7}{*}{140}  &  \multirow{7}{*}{$\mathrm{[Og]} 8s_{1/2}^2 8p_{1/2}^2 5g_{7/2}^8$}  & $7d_{3/2}^2 6f_{5/2}^2 5g_{9/2}^6$ & -93548.8504 & -93320.7516 $\, \bigtriangledown 1$ & -93179.0799 \\
                      &                                                            & $7d_{3/2}^1 6f_{5/2}^3 5g_{9/2}^6$ & -93548.8479 & -93320.7539 $\, \triangle        1$ & -93179.0837 \\
                      &                                                            & $7d_{3/2}^1 6f_{5/2}^4 5g_{9/2}^5$ & -93548.7928 & -93320.6719 $\, \bigtriangledown 1$ & -93178.9986 \\
                      &                                                            & $7d_{3/2}^3 6f_{5/2}^1 5g_{9/2}^6$ & -93548.7738 & -93320.6697 $\, \bigtriangledown 1$ & -93178.9965 \\
                      &                                                            & $6f_{5/2}^4 5g_{9/2}^6           $ & -93548.7689 & -93320.6791 $\, \triangle        2$ & -93179.0103 \\
                      &                                                            & $6f_{5/2}^5 5g_{9/2}^5           $ & -93548.7559 & -93320.6398 & -93178.9679 \\
                      &                                                            & $7d_{3/2}^2 6f_{5/2}^3 5g_{9/2}^5$ & -93548.7488 & -93320.6229 & -93178.9480 \\

\bottomrule
\end{tabular} 
\label{tab:2}
\end{table}
\twocolumngrid\
\par
As one can see from the examples given in Table~\ref{tab:2}, the Breit-interaction correction can change the ground-state configuration within the RAV approximation.
For~$Z=125$, Table~\ref{tab:2} demonstrates that when only the Coulomb interaction is taken into account the ground-state configuration is~$8p^1_{1/2} 6f_{5/2}^4$.
However, when we add the Breit-interaction correction evaluated according to Eq.~(\ref{eq:de_breit}), the configuration~$8p^1_{1/2}  6f_{5/2}^3 5g^1_{7/2}$ turns out to be the lowest-energy one.
All the other six configurations change their order as well.
A similar configuration reordering occurs for the SHE with~$Z=140$ as well. However, in the latter case some configurations retain their positions. 
Without the Breit interaction, the ground-state configuration for~$Z=140$ is~$7d_{3/2}^2 6f_{5/2}^2 5g_{9/2}^6 $, but with this correction taken into account the lowest-energy configuration is~$7d_{3/2}^1 6f_{5/2}^3 5g_{9/2}^6 $.
For both demonstrated cases, the QED correction does not change the ground-state configuration and the sorted configuration list as a whole.
\par
After the configurations with the lowest average energies~$E^{\mathrm{DCBQ}}_{\mathrm{RAV}}(K)$ are found, we explore their level structure using the DCBQ-SRC approach.
To determine the ground-state level, we choose seven configurations with the lowest DCBQ-RAV energies and for each level belonging to these configurations solve the DCBQ-SRC problem~(\ref{eq:e_dcbq_src}).
The configuration with the lowest DCBQ-SRC energy is considered to be the ground-state configuration in the DCBQ-SRC approach, and the corresponding level gives the ground-state level.
Since for most SHEs, a few selected DCBQ-RAV energies are close to each other, we bear in mind that the DCBQ-SRC consideration can change the ground-state configuration. 
\begin{figure}[H]
  \centering
  \includegraphics[width=\linewidth]{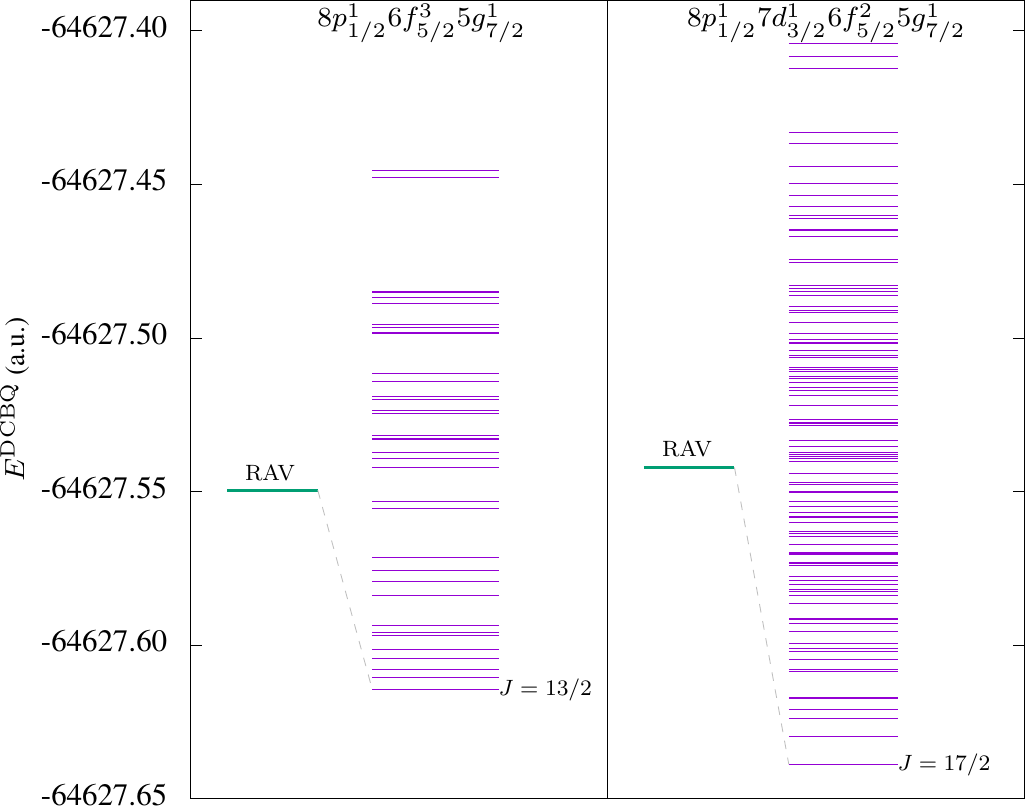}
  \caption{Relativistic-configuration-average energies $E^{\mathrm{DCBQ}}_{\mathrm{RAV}}$ calculated for the configurations~$8p_{1/2}^1 6f_{5/2}^3 5g_{7/2}^1$ (left) and~$8p_{1/2}^1  7d_{3/2}^1 6f_{5/2}^2 5g_{7/2}^1$ (right) of the SHE with~$Z=125$ and all the possible levels which contribute to these average energies.
  For the lowest DCBQ-SRC levels the total angular momentum quantum numbers~$J$ are shown.
  }
  \label{fig:1}
\end{figure}
The last statement is illustrated in Fig.~\ref{fig:1}, where the average energies~$E^{\mathrm{DCBQ}}_{\mathrm{RAV}}(K)$ of the SHE with $Z=125$ are presented for two configurations having the lowest average energies:~$K=8p_{1/2}^1 6f_{5/2}^3 5g_{7/2}^1 $ (left) and~$\tilde{K}=8p_{1/2}^1 7d_{3/2}^1 6f_{5/2}^2 5g_{7/2}^1$ (right).
For each configuration, we show all the levels~$E^{\mathrm{DCBQ}}_{\mathrm{SRC}}(K,J)$ contributing to the relativistic-configuration-average energy.
For the lowest DCBQ-SRC levels, the total angular momenta~$J$ are presented.
One can see, that the average energy of the configuration~$K$ is lower than the energy of the configuration~$\tilde{K}$ by about~$E^{\mathrm{DCBQ}}_{\mathrm{RAV}}(\tilde{K}) -E^{\mathrm{DCBQ}}_{\mathrm{RAV}}(K) = 0.0065 $~a.u.\ (see also Table~\ref{tab:2}).
However, the lowest level~$J=17/2$ of the configuration~$\tilde{K}$ lies lower than the lowest level~$J=13/2$ of the configuration~$K$ by about~$E^{\mathrm{DCBQ}}_{\mathrm{SRC}}(K,13/2) -E^{\mathrm{DCBQ}}_{\mathrm{SRC}}(\tilde{K},17/2) = 0.0245 $~a.u.
This kind of behavior is not a specific feature of the SHE with~$Z=125$, but rather an example of some general trend observed for many other SHEs as well.
\par
The DCBQ-RAV and DCBQ-SRC schemes are one-configuration approaches and therefore they do not take into account the electron-correlation effects i.e. mixing of the different configurations. 
Below, we discuss the influence of the electron-electron correlations on the order of the SHE lowest levels.
To this end, for each $Z$ we perform the independent CI-DFS calculations for seven reference configurations with the lowest DCBQ-RAV energies selected at the previous stage.
For each configuration, we evaluate the three lowest levels and determine their total angular momenta~$J$.
The level with the lowest energy~$E^{\mathrm{DCBQ}}_{\mathrm{CI}}(J)$ is the ground-state level.
Within this approach, the ground-state level depends on how accurately we solve the CI problem.
We investigate the dependence of the level order on the number of virtual orbitals by performing two different CI-DFS calculations referred to as ``DCBQ-CI1'' and ``DCBQ-CI2''.
In both schemes, the single (S) and double (D) excitations from the reference configuration determined at the DCBQ-RAV stage are considered.
The DCBQ-CI1 scheme can be thought of as a CI problem with a relatively small number of the virtual orbitals, whereas the DCBQ-CI2 scheme includes more virtual orbitals.
The list of the active and virtual orbitals used in both CI calculations is presented in Table~\ref{tab:3}.
The occupied orbitals, which are not mentioned in the table, are treated as the frozen core.
The active orbitals as well as the virtual orbitals, which are involved in the most important configurations, are taken as the solutions of the DF equations, whereas the other virtual orbitals are obtained from the solutions of the DFS equations.
\onecolumngrid\
\begin{table}[H]
\centering
\caption{The list of the active and virtual relativistic shells employed in the ``DCBQ-CI1'' and ``DCBQ-CI2'' calculations.
The absence of the lower index indicates that for a given~$l$ both the relativistic orbitals with~$l-1/2$ and~$l+1/2$ are included in the CI problem.
}
\begin{tabular}{
c@{\qquad}
l@{\qquad}
l@{\quad}
l
}

\toprule
\multirow{2}{*}{$Z$} & 
\multirow{2}{*}{Active valence shells} & \multicolumn{2}{c}{Virtual shells} \\
&
&
\multicolumn{1}{c}{DCBQ-CI1} & 
\multicolumn{1}{c}{DCBQ-CI2} \\
\midrule
120 -- 121 & $7p_{3/2} 8s_{1/2} 8p_{1/2} 7d_{3/2}$ & $8p_{3/2} 7d_{5/2}                   $  & $9s_{1/2} 9p 8d 6f 5g                         $  \\
122 -- 123 & $8s_{1/2} 8p_{1/2} 7d_{3/2} 6f_{5/2}$ & $8p_{3/2} 7d_{5/2} 6f_{7/2}          $  & $9s_{1/2} 9p 8d 7f 5g                         $  \\
124 -- 133 & $8p_{1/2} 6f_{5/2} 7d_{3/2} 5g_{7/2}$ & $8p_{3/2} 7d_{5/2} 6f_{7/2} 5g_{9/2} $  & $9s_{1/2} 9p 8d 7f 6g                         $  \\
134 -- 144 & $8p_{1/2} 6f_{5/2} 7d_{3/2} 5g_{9/2}$ & $8p_{3/2} 6f_{7/2} 7d_{5/2}          $  & $9s_{1/2} 9p_{1/2} 8d_{3/2} 7f_{5/2} 6g_{7/2}$  \\
145 -- 146 & $6f_{5/2} 7d_{3/2} 5g_{9/2} 9s_{1/2}$ & $8p_{3/2} 6f_{7/2} 7d_{5/2}          $  & $10s_{1/2} 9p 8d 7f 6g                        $  \\
147 -- 155 & $6f_{5/2} 7d_{3/2} 6f_{7/2} 9s_{1/2}$ & $8p_{3/2} 7d_{5/2}                   $  & $10s_{1/2} 9p 8d 7f 6g                        $  \\
156 -- 160 & $6f_{7/2} 7d_{3/2} 9s_{1/2} 7d_{5/2}$ & $8p_{3/2}                            $  & $10s_{1/2} 9p 8d 7f 6g                        $  \\
161 -- 165 & $7d_{3/2} 7d_{5/2} 9s_{1/2}  8p_{3/2}$ & $                                    $  & $10s_{1/2} 9p 8d 7f 6g                        $  \\
166 -- 168 & $7d_{3/2} 7d_{5/2} 9s_{1/2} 8p_{3/2} 9p_{1/2}$ & $9p_{3/2}                            $  & $10s_{1/2} 10p 8d 7f 6g                       $  \\
169 -- 170 & $7d_{5/2} 9s_{1/2} 8p_{3/2} 9p_{1/2} 7f_{5/2}$ & $7f_{7/2} 9p_{3/2}                   $  & $10s_{1/2} 10p 8d 8f 6g                       $  \\

\bottomrule
\end{tabular} 
\label{tab:3}
\end{table}
\twocolumngrid\
\par
The correlation effects can result in exotic scenarios for the level structure. 
An example is presented in Table~\ref{tab:4}, where the lowest levels with~$J=0,1,2,3$ and the related configurations of the SHE with~$Z=168$ are shown.
One can see that within the DCBQ-SRC approximation, when the electron-electron correlations are neglected, the level with~$J=1$, $E^{\mathrm{DCBQ}}_{\mathrm{SRC}}(J\!=\!1)=-202904.8013$~a.u., lies below the levels  with~$J=2$, $E^{\mathrm{DCBQ}}_{\mathrm{SRC}}(J\!=\!2)=-202904.7871$~a.u.\ and $J=0$, $E^{\mathrm{DCBQ}}_{\mathrm{SRC}}(J\!=\!0)=-202904.7789$~a.u.
If we account for the electronic correlations by means of the DCBQ-CI1 scheme, the level with~$J=1$ rises above the level with~$J=2$ and exceeds it by about~$0.025$~a.u.
When we improve the description of the electron-electron correlations using the DCBQ-CI2 scheme, the level with~$J=1$ becomes the lowest one again and the level with~$J=0$ falls below the level with~$J=2$.
In this case, the difference between the levels with~$J=0$ and~$J=1$ constitutes about~$0.01$~a.u.
This demonstrates that with the improvement of the correlation treatment the changes in the ground-state levels may occur.
More importantly, the dominant configurations of levels with $J=0,1,2$ do not coincide with each other.
Therefore, one can expect that not only levels which belong to the same configuration can interchange, but also the rearrangements involving the levels of other configurations are possible.   
\onecolumngrid\
\begin{table}[H]
\centering
\caption{
The lowest-level energies~$E^{\mathrm{DCBQ}}(J)$ for levels $J=0,1,2,3$ calculated by means of the DCBQ-SRC, DCBQ-CI1, and DCBQ-CI2 schemes for the SHE with $Z=168$ (a.u.).
For the DCBQ-SRC values, the configurations of these levels are presented. For the DCBQ-CI1 and DCBQ-CI2 results, the configurations contributing with the weights of at least $0.05$ are given.
}
\renewcommand{\tabcolsep}{0.1cm}
\begin{tabular}{
l @{\qquad\,}
l
c  @{\qquad\,}
l
c  @{\qquad\,}
l
c}
\toprule
\multirow{2}{*}{$J$} & \multicolumn{2}{c}{DCBQ-SRC} & \multicolumn{2}{c}{DCBQ-CI1} & \multicolumn{2}{c}{DCBQ-CI2} \\ 
                   & \multicolumn{1}{c}{$K$}          & $E^{\mathrm{DCBQ}}_{\mathrm{SRC}}(K,J)$          & \multicolumn{1}{c}{$K$}          & \multicolumn{1}{c}{$E^{\mathrm{DCBQ}}_{\mathrm{CI}}(J)$}          & \multicolumn{1}{c}{$K$}          & \multicolumn{1}{c}{$E^{\mathrm{DCBQ}}_{\mathrm{CI}}(J)$}          \\ 
\midrule                   
\vspace{9pt}
0                   & $9s_{1/2}^2 9p_{1/2}^2$           & $-202904.7789$           & $\!\begin{aligned}[t]
                                                 &0.87 \cdot 9s_{1/2}^2 9p_{1/2}^2 + \\
                                                 &0.07 \cdot 9s_{1/2}^2 8p_{3/2}^2 + \\
                                                 &0.05 \cdot 8p_{3/2}^2 9p_{1/2}^2
                                                 \end{aligned} $& $-202904.8206$           &  $0.91 \cdot 9s_{1/2}^2 9p_{1/2}^2$          &  $-202904.9561$          \\
\vspace{9pt}                                                 
1                   & $9s_{1/2}^2 8p_{3/2}^1 9p_{1/2}^1$           & $-202904.8013$           & $\!\begin{aligned}[t]
                                                 &0.92 \cdot 9s_{1/2}^2 8p_{3/2}^1 9p_{1/2}^1 + \\
                                                 &0.07 \cdot 8p_{3/2}^3 9p_{1/2}^1
                                                 \end{aligned} $& $-202904.8247$           & $0.94 \cdot 9s_{1/2}^2 8p_{3/2}^1 9p_{1/2}^1 $          & $-202904.9652$          \\
\vspace{9pt}                                                 
2                   & $9s_{1/2}^1 8p_{3/2}^2 9p_{1/2}^1$           & $-202904.7871$           & $\!\begin{aligned}[t]
                                                 &0.55 \cdot 9s_{1/2}^1 8p_{3/2}^2 9p_{1/2}^1 + \\
                                                 &0.24 \cdot 9s_{1/2}^1 8p_{3/2}^1 9p_{1/2}^2 + \\
                                                 &0.20 \cdot 9s_{1/2}^1 8p_{3/2}^3
                                                 \end{aligned} $& $-202904.8491$           &  $\!\begin{aligned}[t]
                                                                                &0.81 \cdot 9s_{1/2}^1 8p_{3/2}^2 9p_{1/2}^1 +\\
                                                                                &0.09 \cdot 9s_{1/2}^1 8p_{3/2}^1 9p_{1/2}^2 + \\
                                                                                &0.07 \cdot 9s_{1/2}^1 8p_{3/2}^3
                                                                                \end{aligned}$ & $-202904.9542$ \\
\vspace{9pt}                                                 
3                   & $9s_{1/2}^1 8p_{3/2}^2 9p_{1/2}^1$           & $-202904.7319$           &  $0.99 \cdot 9s_{1/2}^1 8p_{3/2}^2 9p_{1/2}^1$          &  $-202904.7385$          & $0.96 \cdot 9s_{1/2}^1 8p_{3/2}^2 9p_{1/2}^1$           &  $-202904.8835$         \\
\bottomrule
\end{tabular} 
\label{tab:4}
\end{table}
\twocolumngrid\

\onecolumngrid\
\begin{table}[htbp]
\centering
\caption{The ground-state configurations of superheavy elements with atomic numbers $120\leq Z \leq 170$ evaluated within the DCBQ-RAV method, taking into account the Breit interaction and QED effects.
The configurations are shown relative to the closed-shell ones, which are presented in the column ``Closed shells''. 
[Og] corresponds to the configuration of the oganesson atom.
The succeeding records in this column show the relativistic orbitals which have to be added to the previous ones to obtain the closed-shell configurations for heavier atoms.
The results of the present work, ``DCBQ-RAV'', are compared with the results of Refs.~\cite{1977_FrickeB_ADNDT19, 1970_MannJ_JChemPhys, 1996_UmomotoK_JPSJ}.
}
\renewcommand{\tabcolsep}{0.2cm}
\begin{tabular}{
c
c
l
l
l
l
}

\toprule
\multicolumn{1}{c}{$Z$} & \multicolumn{1}{c}{Closed shells} & \multicolumn{1}{c}{DCBQ-RAV} &\multicolumn{1}{c}{Fricke and Soff~\cite{1977_FrickeB_ADNDT19}}  &\multicolumn{1}{c}{Mann and Waber~\cite{1970_MannJ_JChemPhys}} &\multicolumn{1}{c}{Umemoto and Saito~\cite{1996_UmomotoK_JPSJ}}
\\
\midrule

120& [Og] &  $ 8s_{1/2}^2 $       & $ 8s_{1/2}^2 $ &  $ 8s^2 $  &    \\
\midrule
121& +$8s_{1/2}^2$ &  $   8p_{1/2}^1 \,\,            $       & $ 8p_{1/2}^1 $ &  $ 8p^1 $  & $ 8p^1 $   \\
122& &  $  8p_{1/2}^1 \,\, 7d_{3/2}^1 $       & $ 8p_{1/2}^1 \,\, 7d_{3/2}^1 $ &  $  8p^1 7d^1 $  &  $ 8p^2 $  \\
123& &  $  8p_{1/2}^1 \,\, 7d_{3/2}^1 \,\, 6f_{5/2}^1 $       & $ 8p_{1/2}^1 \,\, 7d_{3/2}^1 \,\, 6f_{5/2}^1 $ & $  8p^1 7d^1 6f^1 $   &  $ 8p^1 7d^1 6f^1 $  \\
124& &  $ 8p_{1/2}^1 \,\, 6f_{5/2}^3 $       & $ 8p_{1/2}^1 \,\, 6f_{5/2}^3 $ &  $ 8p^1 6f^3 $  &  $8p^2 6f^2 $   \\
125& &  $ 8p_{1/2}^1 \,\, 6f_{5/2}^3 \,\, 5g_{7/2}^1 $       & $ 8p_{1/2}^1 \,\, 6f_{5/2}^3 \,\, 5g_{7/2}^1 $ & $ 8p^1 6f^3 5g^1 $   &  $8p^1 6f^4 $   \\
126& &  $ 8p_{1/2}^2 \,\, 6f_{5/2}^2 \,\, 5g_{7/2}^2 $       & $ 8p_{1/2}^1 \,\, 7d_{3/2}^1 \,\, 6f_{5/2}^2 \,\, 5g_{7/2}^2 $ & $ 8p^2 6f^2 5g^2 $   &  $ 8p^1 6f^4 5g^1 $  \\
\midrule
127& +$8p_{1/2}^2$ &  $ 6f_{5/2}^2 \,\, 5g_{7/2}^3 \,\,            $       & $  6f_{5/2}^2 \,\, 5g_{7/2}^3 $ &  $ 8p^2 6f^2 5g^3  $  &  $ 8p^2 6f^3 5g^2 $  \\
128& &  $ 6f_{5/2}^2 \,\, 5g_{7/2}^4 \,\,            $       & $  6f_{5/2}^2 \,\, 5g_{7/2}^4 $ & $ 8p^2 6f^2 5g^4  $   &  $8p^2 6f^3 5g^3 $\\  
129& &  $  6f_{5/2}^2 \,\, 5g_{7/2}^5 \,\,            $       & $  6f_{5/2}^2 \,\, 5g_{7/2}^5 $ & $ 8p^2 6f^2 5g^5  $    &  $8p^2 6f^3 5g^4 $ \\ 
130& &  $ 6f_{5/2}^2 \,\, 5g_{7/2}^6 \,\,            $       & $  6f_{5/2}^2 \,\, 5g_{7/2}^6 $ & $ 8p^2 6f^2 5g^6  $    &   $8p^2 6f^3 5g^5 $ \\
131& &  $  6f_{5/2}^2 \,\, 5g_{7/2}^7 $       & $  6f_{5/2}^2 \,\, 5g_{7/2}^7 $ & $ 8p^2 6f^2 5g^7  $    &  $8p^2 6f^3 5g^6 $ \\
132& &  $ 7d_{3/2}^1 \,\, 6f_{5/2}^1 \,\, 5g_{7/2}^8 $       & $ 6f_{5/2}^2 \,\, 5g_{7/2}^8$&    &    \\
\midrule
133& +$5g_{7/2}^8$ &  $ 6f_{5/2}^3 \,\,             $       & $ 6f_{5/2}^3 $ &    &    \\
134& &  $ 6f_{5/2}^4 \,\,            \,\,            $       & $ 6f_{5/2}^4 $&    &    \\
135& &  $ 6f_{5/2}^4 \,\, 5g_{9/2}^1 \,\,            $       & $ 6f_{5/2}^4 \,\, 5g_{9/2}^1 $&    &    \\
136& &  $ 6f_{5/2}^4 \,\, 5g_{9/2}^2 $       & $ 6f_{5/2}^4 \,\, 5g_{9/2}^2 $ &    &    \\
137& &  $ 7d_{3/2}^1 \,\, 6f_{5/2}^3 \,\, 5g_{9/2}^3 $       & $ 7d_{3/2}^1 \,\, 6f_{5/2}^3 \,\, 5g_{9/2}^{3} $ &    &    \\
138& &  $ 7d_{3/2}^1 \,\, 6f_{5/2}^3 \,\, 5g_{9/2}^4 $       & $ 7d_{3/2}^1 \,\, 6f_{5/2}^3 \,\, 5g_{9/2}^{4} $&    &    \\
139& &  $ 7d_{3/2}^1 \,\, 6f_{5/2}^3 \,\, 5g_{9/2}^5 $       & $ 7d_{3/2}^2 \,\, 6f_{5/2}^2 \,\, 5g_{9/2}^{5} $ &    &    \\
140& &  $ 7d_{3/2}^1 \,\, 6f_{5/2}^3 \,\, 5g_{9/2}^6 $       & $ 7d_{3/2}^1 \,\, 6f_{5/2}^3 \,\, 5g_{9/2}^{6} $&    &    \\
141& &  $ 7d_{3/2}^2 \,\, 6f_{5/2}^2 \,\, 5g_{9/2}^7 $       & $ 7d_{3/2}^2 \,\, 6f_{5/2}^2 \,\, 5g_{9/2}^{7} $&    &    \\
142& &  $ 7d_{3/2}^2 \,\, 6f_{5/2}^2 \,\, 5g_{9/2}^8 $       & $ 7d_{3/2}^2 \,\, 6f_{5/2}^2 \,\, 5g_{9/2}^{8} $ &    &    \\
143& &  $ 7d_{3/2}^2 \,\, 6f_{5/2}^2 \,\, 5g_{9/2}^9 $       & $ 7d_{3/2}^2 \,\, 6f_{5/2}^2 \,\, 5g_{9/2}^{9} $ &    &    \\
144& &  $ 7d_{3/2}^3 \,\, 6f_{5/2}^1 \,\, 5g_{9/2}^{10} \,\,            $       & $ 7d_{3/2}^3 \,\, 6f_{5/2}^1 \,\, 5g_{9/2}^{10} $&    &    \\
\midrule
145& +$5g_{9/2}^{10}$ &  $ 7d_{3/2}^2 \,\, 6f_{5/2}^3 $       & $ 7d_{3/2}^2 \,\, 6f_{5/2}^3 $ &    &    \\
146& &  $ 7d_{3/2}^2 \,\, 6f_{5/2}^4 $      & $ 7d_{3/2}^2 \,\, 6f_{5/2}^4 $ &    &    \\
147& &  $ 7d_{3/2}^2 \,\, 6f_{5/2}^5 \,\,            \,\,            $       & $ 7d_{3/2}^2 \,\, 6f_{5/2}^5 $ &    &    \\
148& &  $ 7d_{3/2}^2 \,\, 6f_{5/2}^6 $       & $ 7d_{3/2}^2 \,\, 6f_{5/2}^6 $ &    &    \\
\midrule
149& +$6f_{5/2}^6$ &  $ 7d_{3/2}^3 \,\,        \,\,            $       & $ 7d_{3/2}^3 $&    &    \\
150& &  $ 7d_{3/2}^4 \,\,            \,\,            $       & $ 7d_{3/2}^4 $&    &    \\
151& &  $ 7d_{3/2}^3 \,\,  6f_{7/2}^2 \,\,            $       & $ 7d_{3/2}^3 \,\, 6f_{7/2}^2 $ &    &    \\
152& &  $ 7d_{3/2}^3 \,\,  6f_{7/2}^3 \,\,            $       & $ 7d_{3/2}^3 \,\, 6f_{7/2}^3 $&    &    \\
153& &  $ 7d_{3/2}^2 \,\,  6f_{7/2}^5 \,\,            $       & $ 7d_{3/2}^2 \,\, 6f_{7/2}^5 $ &    &    \\
154& &  $ 7d_{3/2}^2 \,\, 6f_{7/2}^6 \,\,            $       & $ 7d_{3/2}^2 \,\, 6f_{7/2}^6 $&    &    \\
155& &  $ 9s_{1/2}^1 \,\, 6f_{7/2}^8  $       & $ 7d_{3/2}^2 \,\, 6f_{7/2}^7 $  &    &    \\
\midrule
156& +$6f_{7/2}^8$ &  $ 7d_{3/2}^2 $       & $ 7d_{3/2}^2 $ &    &    \\
157& &  $ 7d_{3/2}^3   \,\,            \,\,            $       & $ 7d_{3/2}^3  $&    &    \\
158& &  $ 7d_{3/2}^4  $       & $ 7d_{3/2}^4  $ &    &    \\
\midrule
159& +$7d_{3/2}^4$ &  $   9s_{1/2}^1  $       & $   9s_{1/2}^1  $ &    &    \\
160& &  $  7d_{5/2}^1 \,\, 9s_{1/2}^1 $       & $  7d_{5/2}^1 \,\, 9s_{1/2}^1 $ &    &    \\
161& &  $  7d_{5/2}^2 \,\, 9s_{1/2}^1 $       & $  7d_{5/2}^2 \,\, 9s_{1/2}^1 $ &    &    \\
162& &  $  7d_{5/2}^4  $       & $ 7d_{5/2}^4  $ &    &    \\
163& &  $  7d_{5/2}^5  $       & $ 7d_{5/2}^5  $ &    &    \\
164& &  $  7d_{5/2}^6  $       & $ 7d_{5/2}^6  $ &    &    \\
\midrule
165& +$7d_{5/2}^6$ &  $   9s_{1/2}^1 \,\,            $       & $ 9s_{1/2}^1 $ &    &    \\
166& &  $ 9s_{1/2}^2 $       & $ 9s_{1/2}^2 $ &    &    \\
\midrule
167& +$9s_{1/2}^2$ &  $ 8p_{3/2}^1 \,\,  \,\,            $       &  $9p_{1/2}^1 $&    &    \\
168& &  $ 8p_{3/2}^1 \,\,  9p_{1/2}^1 $       & $ 9p_{1/2}^2 $ &    &    \\
169& &  $ 8p_{3/2}^1 \,\,  9p_{1/2}^2 $       & $ 8p_{3/2}^1 \,\,  9p_{1/2}^2 $ &    &    \\
170& &  $ 8p_{3/2}^2 \,\, 9p_{1/2}^2 $       & $ 8p_{3/2}^2 \,\, 9p_{1/2}^2 $ &    &    \\

\bottomrule

\end{tabular}
\label{tab:5}
\end{table}
\twocolumngrid
\section{Results}\label{sec:results}
To begin with, we calculate the average energies of the configurations including the Breit interaction and QED effects within the DCBQ-RAV method.
For all the SHEs in the range~$120\leq Z \leq 170$, the configurations with the lowest average energy are presented in Table~\ref{tab:5}.
The ground-state configurations are shown relative to the closed-shell configurations given in the second column.
Our results (the column ``DCBQ-RAV'') are compared with the results of Ref.~\cite{1977_FrickeB_ADNDT19}, where the calculations were performed using the Dirac-Fock-Slater method for~$Z$ up to~$173$, the results of Ref.~\cite{1970_MannJ_JChemPhys}, where the DF method was employed for~$Z$ up to~$131$, and, finally, the results of Ref.~\cite{1996_UmomotoK_JPSJ}, where the calculations were based on the relativistic density functional theory and $121\leq Z\leq 131$ were considered.
We note that in Ref.~\cite{1970_MannJ_JChemPhys} the Gaunt-interaction correction was taken into account perturbatively, while the Coulomb electron-electron interaction was treated self-consistently within the density functional theory. 
Moreover, since the non-relativistic notations were used in Refs.~\cite{1970_MannJ_JChemPhys, 1996_UmomotoK_JPSJ}, we retain them in Table~\ref{tab:5} without any changes.

\par
Within the DCBQ-RAV approximation, the general trend of the occupation rule with the growth of $Z$ is as follows: after the $8s_{1/2}$~shell is filled for~$Z=120$, the electrons begin to occupy the $8p_{1/2}$,~$7d_{3/2}$, and~$6f_{5/2}$~shells.
According to our calculations, the first electron in the $5g_{7/2}$~shell appears for~$Z=125$. Starting from this atomic number, the~$5g$ series begins.
For $Z=126$, the $8p_{1/2}$~shell becomes the closed one.
With a further increase for~$Z$, the $5g_{7/2}$ and~$5g_{9/2}$~shells are subsequently occupied with the electrons.
The partially occupied $7d_{3/2}$ and~$6f_{5/2}$~shells remain the valence ones and their occupation numbers exhibit little changes.
This $5g$-occupation process is completed at~$Z=144$ which has the fully occupied $5g_{7/2}$ and~$5g_{9/2}$~shells and partially occupied $7d_{3/2}$ and~$6f_{5/2}$~shells.
\par
After both the relativistic $5g$~shells are filled, the $6f$~shells begin to be occupied quite systematically:~$6f_{5/2}$ and~$6f_{7/2}$ become fully occupied for~$Z=148$ and $Z=155$, respectively.
The $7d_{3/2}$~shell is unoccupied for~$Z=155$ and it remains partially occupied for the other $Z$ in this region becoming closed only for~$Z=158$.
Finally, the SHE with~$Z=164$ has the configuration~$\mathrm{[Og]} 8s_{1/2}^2 8p_{1/2}^2 5g_{7/2}^8 5g_{9/2}^{10} 6f_{5/2}^6 6f_{7/2}^8 7d_{3/2}^4 7d_{5/2}^6$ with all the relativistic shells being occupied.
\par
Notably, the $8p_{3/2}$~shell is not filled along the sequence~$Z=120-166$ that is due to a large spin-orbital splitting of the $8p$~shell.
The DF-RAV calculations for~$Z=166$ shows that the spin-orbital splitting of the $8p$~shell is about~$80$~eV.
The first~$8p_{3/2}$ electron appears in the SHE with~$Z=167$, after the $9s_{1/2}$~shell becomes the closed one.
However, for~$Z=168$ the $9p_{1/2}$~shell turns out to be more energetically advantageous than the~$8p_{3/2}$ one.
The $8p_{3/2}$~shell is not fully occupied even for the last considered SHE with~$Z=170$.
Another remarkable observation found in our DCBQ-RAV calculations is that for~$Z=155$ the configuration with the valence $9s_{1/2}$~electron turn out to be more energetically beneficial than the configuration with $7d_{3/2}$~electrons, whereas its neighbors ---~$Z=154$ and~$Z=156$ --- have two electrons in the $7d_{3/2}$~shell.
Next time the $9s_{1/2}$~electron appears in the series $Z=159$\,--\,$161$, and, finally, the $9s_{1/2}$~shell establishes on the regular basis starting from the element with~$Z=165$.
\par
Throughout the calculations we found that the ground-state configuration may change due to the Breit-interaction corrections, see the discussion in Sec.~\ref{sec:details}.
This kind of changes is observed for~$Z=125$ and~$Z=140$ and never occurs for the other values of $Z$.
Concerning the QED corrections, we deduce that within the DCBQ-RAV approach they never change the ground-state configuration for the SHEs under consideration.
In general, our DCBQ-RAV ground-state configurations coincide with the DF-RAV ones, obtained without the Breit and QED corrections, in all cases except for~$Z=125$ and~$Z=140$. 
\par
Our DCBQ-RAV results for~$Z=120$\,--\,$131$ are in full agreement with the results of Ref.~\cite{1970_MannJ_JChemPhys}.
The obtained ground-state configurations agree with the related results of Ref.~\cite{1996_UmomotoK_JPSJ} for the SHEs with~$121 \leq Z \leq 123$, but differ for the other available values of~$Z$.
Our DF-RAV results differ from the results of Ref.~\cite{1977_FrickeB_ADNDT19} obtained without the Breit and QED corrections for eight of the considered SHEs, namely, for~$Z=125, 126, 132, 139, 140, 155, 167$, and $168$.
For~$Z=155$, our results predict that $9s_{1/2}$ electron unexpectedly jumps in the~$6f_{7/2}$-occupation sequence.
Perhaps the configuration with the valence $9s_{1/2}$~electron was not considered in Ref.~\cite{1977_FrickeB_ADNDT19}. 
As for the other discrepancies, they seem to have a non-systematical nature and might be due to the Slater exchange-interaction approximation used in Ref.~\cite{1977_FrickeB_ADNDT19}. Nevertheless, the real reasons for these deviations remain unclear to us.
\par
Proceeding with the analysis, we are aimed at finding the configuration of the lowest-energy level.
We employ the CI-DFS method using the one-configuration scheme DCBQ-SRC as well as the more elaborated schemes DCBQ-CI1 and DCBQ-CI2.
As in the DCBQ-RAV approach, in these schemes the Breit and QED corrections are included, however, in the non-perturbative manner.
The thorough description of the CI calculations is presented in Sec.~\ref{sec:details}.
\par
In Table~\ref{tab:6}, we give the levels with the lowest DCBQ energies for the SHEs with~$120 \leq Z \leq 170$ obtained in three considered CI schemes.
Additionally, the quantum numbers~$J$ of these levels are listed.
For the DCBQ-SRC results, the configurations which the ground-state levels belong to are given.
The DCBQ-CI1 and DCBQ-CI2 results include the electron-electron correlation effects.
For these data, we list the configurations contributing to the ground levels with the weight of at least~$0.05$.
Following the structure of the previous table, the configurations are given relative to the closed-shell ones.
The obtained results are compared with the results of the previous multiconfiguration Dirac-Fock calculations~\cite{2006_NefedovV_DoklPhysChem}. 
The non-relativistic notations of Ref.~\cite{2006_NefedovV_DoklPhysChem} are retained.
\par
A comparison of Tables~\ref{tab:5} and~\ref{tab:6}  shows that the configurations of the ground levels obtained within the SRC approach differ from  the RAV ground-state configurations in almost half of the cases (for convenience, the corresponding values of $Z$ are typed in a bold font). This result indicates the complex level structure of the SHEs, which is discussed in details for $Z=125$ in Sec. \ref{sec:details}.

\par
The subsequent discussion consists of two parts. At first, we identify general trends for the results of the many-configuration calculations and compare them with the single-configuration ones. 
We note, that the DCBQ-CI1 and DCBQ-CI2 results are, in general, not much different. Therefore, in this part we often drop the indices "1" or "2" and use the generalized designation "DCBQ-CI" for the many-configuration calculations. In the second part, we compare the results obtained by the configuration-interaction method CI1 and CI2 with each other. 
\par
Exactly as the DCBQ-RAV scheme predicts, our many-configuration DCBQ-CI results detect the first appearance of the~$5g$ electron in the ground state for the SHE with~$Z=125$.
However, in contrast to the DCBQ-RAV results, the DCBQ-CI schemes predict that the $5g$ shell becomes closed for $Z=145$ instead of $Z=144$. 
In the range~$Z=125$\,--\,$132$, the many-configuration calculations reveal that the dominant configurations of the obtained ground-state levels in all eight cases differ from the ones obtained within the DCBQ-RAV approach.
Moreover, a configuration mixing in the ground states takes place for some SHEs in range~$Z=125$\,--\,$145$ as the $5g$ shells are gradually occupied. 
In most of the considered cases, the weights of the dominant configurations lie in range~$0.80$\,--\,$0.90$.
The configurations with the different occupation numbers for the $8p_{1/2}$, 
$7d_{3/2}$, $6f_{5/2}$, and $5g_{7/2,9/2}$ shells are admixed.
The DCBQ-CI schemes show that starting from~$Z=130$ the dominant configuration of the ground-state level has the $8p_{1/2}$~shell fully occupied.
However, the configurations with the partially occupied $8p_{1/2}$~shell contribute (with the weights about~$0.05$ or higher) to these levels up approximately ~$Z\approx135-137$.
\par
A mixture of the configurations with the partially occupied~$7d_{3/2}$ and~$6f_{5/2}$ shells occurs also in the range~$Z=147$\,--\,$151$.
The situation with the ground states becomes more clear starting from the SHE with~$Z=152$, when the $6f_{5/2}$~shell turns out to be fully occupied.
Up to~$Z=165$, the weights of the dominant configurations are larger than~$0.90$, and in most of the cases the dominant configurations of the ground-state levels coincide with the ground-state configurations obtained within the DCBQ-RAV approach.
In particular, the fact that the SHE with $Z=164$ possesses the ground-state configuration with all the relativistic shells closed is confirmed by the more elaborated methods.
The SHEs with~$Z=168$ and~$Z=169$ demonstrate within the DCBQ-CI1 scheme poorly resolved dominant configurations of the ground-state levels.
For instance, the DCBQ-CI1 weight of the dominant configuration for~$Z=168$ is only~$0.55$, which was not the case even for the SHEs with the open $5g_{7/2}$ and~$5g_{9/2}$~shells. 
However, increasing the number of the active orbitals remedies the situation, and for~$Z=168$ the DCBQ-CI2 scheme yields the dominant-configuration weight equal to~$0.92$.
This is due to the fact that the levels interchange, see the corresponding discussion in Sec.~\ref{sec:details}.
\par
The overall trends obtained in our many-configuration calculations are the following. 
First, the configurations which have the lowest levels within the DCBQ-SRC approach are the dominant ones contributing to the ground-state levels within the DCBQ-CI approach in about~$80\%$ of the considered cases.
Second, the ground-state levels obtained without the electronic correlations using the DCBQ-SRC scheme in about~$75\%$ of the cases coincide with the ones obtained by means of the DCBQ-CI approach.
The deviations are mainly concentrated in the range~$Z=131-138$, where the $5g_{7/2}$ and~$5g_{9/2}$~shells are partially occupied and strong interaction between several configurations takes place.
The simultaneous change of the dominant configuration and the ground-state level when passing from the DCBQ-SRC to the DCBQ-CI method occurs for, e.g.,~$Z=131$. In this case the first scheme yields~$J_{\mathrm{SRC}}=25/2$ of the configuration~$K_{\mathrm{SRC}}=8p_{1/2}^1 7d_{3/2}^1 6f_{5/2}^3 5g_{7/2}^6$,
whereas the second scheme predicts the lowest level to be~$J_{\mathrm{CI}}=21/2$ with the dominant configuration being~$K_{\mathrm{CI}}=8p_{1/2}^2 6f_{5/2}^3 5g_{7/2}^6$ with the weight of about~$0.82-0.85$.
\par
Now we proceed to contrast of the two DCBQ-CI schemes results. Compared to the DCBQ-CI1 data, the more accurate treatment of the electron-electron correlations by means of the DCBQ-CI2 approach results in the changes of the ground-state level in 4 of 51 cases.
In 3 of these 4 cases, the configuration which gives the maximum contribution to the ground-state level changes as well.
These SHEs, which need particular attention, are the ones with~$Z=130,137,143$, and~$168$.
For instance, for~$Z=130$, the level~$J=14$ of the dominant configuration~$K=8p_{1/2}^1 7d_{3/2}^1 6f_{5/2}^3 5g_{7/2}^5$ is predicted to be the ground-state one in both DCBQ-SRC and DCBQ-CI1 schemes: $K_{\mathrm{SRC}}=K_{\mathrm{CI1}}=K$.
However, the electronic correlations evaluated by means of the DCBQ-CI2 scheme change the ground-state level, and it becomes~$J=12$ with the dominant configuration being~$K_{\mathrm{CI2}}=8p_{1/2}^2 6f_{5/2}^3 5g_{7/2}^5\ne K$.
\par
We compare our DCBQ-CI2 results with the only available systematic many-configuration calculations of Ref.~\cite{2006_NefedovV_DoklPhysChem} where the Breit interaction was taken into account as well. 
Since the quantum numbers~$J$ which characterize the ground-state levels are not presented in that paper, we are able to compare only the configurations.
We found a disagreement in the configurations contributing to the ground states for the SHEs with~$Z=123$\,--\,$128$, $Z=130$, $Z=136$\,--\,$137$, $Z=143$\,--\,$144$, $Z=152$\,--\,$156$, and $Z=163$.
It is difficult to reveal a possible reason of the discrepancy due to the lack of the computational details given in Ref.~\cite{2006_NefedovV_DoklPhysChem}.
\par
The changes of the ground-state levels in transition from the DCBQ-CI1 to the DCBQ-CI2 calculations and the deviations from the previous results raise the following question: can hypothetical larger CI calculations change the obtained ground states as the DCBQ-CI2 scheme changes the ground states in comparison with the DCBQ-CI1 one?
A comprehensive answer can be given only within the scope of the corresponding large-scale CI calculations.
However, to get an idea of the cases for which the correlation effects may change the dominant configuration of the ground-state level, we investigate the behavior of the energy difference between the ground-state level and the closest level belonging to a different dominant configuration for both our DCBQ-CI calculations.
This study allows us to determine whether the ground-state level is in some sense isolated from levels of other configurations and whether the electronic correlations break down this isolation.
The absolute values of the corresponding differences are presented in Table~\ref{tab:7}.
The SHE with $Z=120$ is omitted in Table~\ref{tab:7}, since it possesses the ground-state configuration $K^{\ast}=[\mathrm{Og}]8s_{1/2}^2$ that causes no doubt.
\setcounter{table}{6}
\onecolumngrid\
\begin{table}[hbtp]
\centering
\caption{The absolute values of the energy difference between the ground-state level of the dominant configuration $K^{\ast}$ and the closest excited level belonging to the dominant configuration which is different from $K^{\ast}$ for the SHEs in the range
$121\leq Z \leq170$ (a.u.). 
The results are presented for the DCBQ-CI1 and DCBQ-CI2 schemes.}
\begin{tabular}{
S[table-format=3] 
S[table-format=2.4]
S[table-format=2.6]
S[table-format=3]
S[table-format=2.4]
S[table-format=2.6]
S[table-format=3]
S[table-format=2.4]
S[table-format=2.6]
S[table-format=3]
S[table-format=2.4]
S[table-format=2.6]
S[table-format=3]
S[table-format=2.4]
S[table-format=2.5]
}

\toprule
\multicolumn{1}{c}{$Z$} & \multicolumn{1}{c}{CI1} & \multicolumn{1}{c}{CI2} & \multicolumn{1}{c}{$Z$} & \multicolumn{1}{c}{CI1} & \multicolumn{1}{c}{CI2} & \multicolumn{1}{c}{$Z$} & \multicolumn{1}{c}{CI1} & \multicolumn{1}{c}{CI2} & \multicolumn{1}{c}{$Z$} & \multicolumn{1}{c}{CI1} & \multicolumn{1}{c}{CI2} & \multicolumn{1}{c}{$Z$} & \multicolumn{1}{c}{CI1} & \multicolumn{1}{c}{CI2}\\
\midrule
121 & 0.0399 & 0.0378 & 131  & 0.0047 & 0.0069 & 141 & 0.0195 & 0.0106 & 151 & 0.0176 & 0.0114 & 161 & 0.0103 & 0.0059 \\
122 & 0.0127 & 0.0107 & 132  & 0.0083 & 0.0118 & 142 & 0.0143 & 0.0046 & 152 & 0.0059 & 0.0122 & 162 & 0.0094 & 0.0016 \\
123 & 0.0299 & 0.0351 & 133  & 0.0254 & 0.0204 & 143 & 0.0015 & 0.0116 & 153 & 0.0062 & 0.0129 & 163 & 0.0111 & 0.0176 \\
124 & 0.0050 & 0.0046 & 134  & 0.0313 & 0.0270 & 144 & 0.0013 & 0.0129 & 154 & 0.0107 & 0.0008 & 164 & 0.0492 & 0.0542 \\
125 & 0.0062 & 0.0057 & 135  & 0.0165 & 0.0125 & 145 & 0.0539 & 0.0376 & 155 & 0.0108 & 0.0257 & 165 & 0.0424 & 0.0443 \\
126 & 0.0106 & 0.0126 & 136  & 0.0062 & 0.0095 & 146 & 0.0352 & 0.0273 & 156 & 0.0136 & 0.0110 & 166 & 0.0153 & 0.0305 \\
127 & 0.0070 & 0.0089 & 137  & 0.0011 & 0.0043 & 147 & 0.0398 & 0.0349 & 157 & 0.0208 & 0.0212 & 167 & 0.0022 & 0.0096 \\
128 & 0.0116 & 0.0123 & 138  & 0.0248 & 0.0303 & 148 & 0.0715 & 0.0760 & 158 & 0.0529 & 0.0555 & 168 & 0.0244 & 0.0091 \\
129 & 0.0071 & 0.0034 & 139  & 0.0370 & 0.0322 & 149 & 0.0583 & 0.0568 & 159 & 0.0568 & 0.0581 & 169 & 0.0066 & 0.0087 \\
130 & 0.0022 & 0.0009 & 140  & 0.0237 & 0.0088 & 150 & 0.0295 & 0.0300 & 160 & 0.0308 & 0.0285 & 170 & 0.0287 & 0.0324 \\
\bottomrule

\end{tabular}
\label{tab:7}
\end{table}
\twocolumngrid\
\par
As it is seen from Table~\ref{tab:7}, some SHEs have a clear separation of the ground-state level from levels of other configurations which almost does not depend on the correlation-treatment scheme.
For instance, for~$Z=121$, the separations of the levels in the DCBQ-CI1 and DCBQ-CI2 schemes constitute~$0.0399$~a.u.\ and~$0.0378$~a.u., respectively.
In other cases, the ground-state level becomes more isolated from the levels of other configurations as the correlation treatment is improved.
So, for $Z=155$, the separation increases from~$0.0108$~a.u.\ in the DCBQ-CI1 scheme to~$0.0257$~a.u.\ in the DCBQ-CI2 one.
In spite of this, it is difficult to formulate for all the systems under consideration 
a reliable criteria to determine if the dominant configuration 
contributing to the ground-state level does change with increase of the configuration-space.
From this point of view, the most suspicious SHEs are the ones which have a small (less than a few thousandths of a.u.) separation between the considered levels within the DCBQ-CI2 scheme.
In addition, we also include in this category the cases where the separation between the levels significantly decreases in the DCBQ-CI2 scheme compared to the DCBQ-CI1 results.
Analyzing the data in Table~\ref{tab:7}, we consider the SHEs with~$Z=129, 130,137,140,142,154,161,162, 168$, and $169$ as those that can possibly have a different dominant configuration of the ground-state level than the one obtained within the DCBQ-CI2 scheme.
These elements have to be studied within the more elaborated electron-correlation calculations.

\section{Conclusion}\label{sec:conclusion}
In the scope of the present paper, we have performed the extensive relativistic study of the ground states of the superheavy elements in the range~$120 \leq  Z \leq 170$.
The Breit interaction is rigorously taken into in the calculations, and the QED effects are considered within the model-QED-operator approach~\cite{2013_ShabaevV_PhysRevA, 2015_ShabaevV_CompPhysComm, *2018_ShabaevV_CompPhysComm, 2022_MalyshevA_PRA106}.
The ground-state configurations are first determined by means of the Dirac-Fock method in the relativistic-configuration-average approximation. 
It is deduced that the QED effects can not change the ground-state configuration in contrast to the Breit interaction.
\par
To resolve the level structure of the configurations, the ground-state levels are found using the configuration-interaction method in the basis of the Dirac-Fock-Sturm orbitals.
We study the general trends in the order of occupation of orbitals in the SHE.
We obtain that in spite of the complex electronic structure of the considered SHEs, the ground-state levels have distinct dominant configurations with the weights exceeding~$0.85$.
Finally, we demonstrate that the electron-correlation effects can change the dominant configuration of the ground-state level.
For some SHEs, the large-scale calculations are needed in order to more reliably determine the ground states and the structure of low-lying energy levels.
Nevertheless, the ground-state configurations of the SHEs obtained in the present work within the many-configuration approach can be used as a solid basis for accurate calculations of various atomic properties of these elements as well as to examine the role of the electron-electron correlations, QED, and relativistic effects on the Periodic Law.

\section{Acknowledgements}
We thank Yu.~Ts.~Oganessian for stimulating discussions. 
Valuable conversations with E.~Eliav, V.~Pershina, and A.~V.~Titov are also gratefully 
acknowledged.
The work was supported by the Ministry of Science and Higher Education of the Russian Federation within~Grant~No.~075-10-2020-117.

\newpage
\bibliographystyle{apsrev4-2}
\bibliography{main}

\clearpage

\setcounter{table}{5}
\onecolumngrid\
\LTcapwidth=\textwidth
\begin{longtable}{ 
>{\footnotesize}c
>{\footnotesize}c @{\quad} 
>{\footnotesize}l
>{\footnotesize}c @{\quad\,} 
>{\footnotesize}l
>{\footnotesize}c @{\quad\,} 
>{\footnotesize}l
>{\footnotesize}c @{\quad} 
>{\footnotesize}l
}
\caption{The levels with lowest total energies, the main configurations contributing to them, and the total angular momenta ~$J$ evaluated by means of the DCBQ-SRC, DCBQ-CI1, and DCBQ-CI2 schemes for the SHEs with the atomic numbers~$120\leq Z \leq 170$.
For the DCBQ-CI1 and DCBQ-CI2 results, the configurations with weights of at least~$0.05$ are presented.
The configurations obtained are given relative to the closed-shell configurations listed in the column ``Closed shells''.
In the first column, the values of $Z$ typed in the bold font indicate the cases when the ground-state configurations obtained within the DCBQ-RAV and DCBQ-SRC methods differ.  
In addition, the following notations around~$Z$ are adopted to assist the reader in navigation through the table.
The underline ``$\underline{\phantom{10}}$'' means that the ground-state $J_{\mathrm{SRC}}$~level evaluated using the DCBQ-SRC approach differs from the~$J_{\mathrm{CI1}}$ one calculated by means of the DCBQ-CI1 approach, $J_{\mathrm{SRC}} \neq J_{\mathrm{CI1}}$. The left vertical line ``$\lvert{\phantom{10}}$'' signalizes that the configuration~$K_{\mathrm{SRC}}$ which the ground-state SRC level belongs to differs from the dominant configuration~$K_{\mathrm{CI1}}$ of the ground-state CI1 level, $K_{\mathrm{SRC}} \neq K_{\mathrm{CI1}}$.
The overline ``$\overline{\phantom{10}}$'' represents the fact that $J_{\mathrm{CI1}} \neq J_{\mathrm{CI2}}$. Finally, the right vertical line ``${\phantom{10}\rvert}$'' stands for the case $K_{\mathrm{CI1}} \neq K_{\mathrm{CI2}}$.
The obtained ground-state levels are compared with the results of Ref.~\cite{2006_NefedovV_DoklPhysChem}.
The non-relativistic notation of Ref.~\cite{2006_NefedovV_DoklPhysChem} are retained.
}
\\
\toprule
\vspace{3pt}
\multirow{2}{*}{$Z$} &
\multirow{2}{*}{Closed} &
\multirow{2}{*}{DCBQ-SRC} &
\multirow{2}{*}{$J_{\mathrm{SRC}}$} &
\multirow{2}{*}{DCBQ-CI1} &
\multirow{2}{*}{$J_{\mathrm{CI1}}$} &
\multirow{2}{*}{DCBQ-CI2} &
\multirow{2}{*}{$J_{\mathrm{CI2}}$} &
\multirow{2}{*}{Ref.~\cite{2006_NefedovV_DoklPhysChem}} \\
&
Shells &
&
&
&
&
&
&
\\
\midrule
\endfirsthead

\multicolumn{9}{c}{{\tablename\ \thetable{}. \textit{(Continuation.)}}}
\\
\toprule
\vspace{3pt}
\multirow{2}{*}{$Z$} &
\multirow{2}{*}{Closed} &
\multirow{2}{*}{DCBQ-SRC} &
\multirow{2}{*}{$J_{\mathrm{SRC}}$} &
\multirow{2}{*}{DCBQ-CI1} &
\multirow{2}{*}{$J_{\mathrm{CI1}}$} &
\multirow{2}{*}{DCBQ-CI2} &
\multirow{2}{*}{$J_{\mathrm{CI2}}$} &
\multirow{2}{*}{Ref.~\cite{2006_NefedovV_DoklPhysChem}} \\
&
Shells &
&
&
&
&
&
&
\\
\midrule
\endhead

\vspace{9pt} 
120 &      [Og]             & $8s_{1/2}^2                                   $ &  $0   $   & $0.94 \cdot 8s_{1/2}^2                                          $       &$0   $   & $0.93 \cdot 8s_{1/2}^2$             & $0   $  & $8s^2$  \\
\vspace{9pt} 
121 &                       & $8s_{1/2}^2 8p_{1/2}^1                        $ &  $1/2 $   & $0.92 \cdot 8s_{1/2}^2 8p_{1/2}^1                                          $       &$1/2 $   & $0.91 \cdot 8s_{1/2}^2 8p_{1/2}^1$  & $1/2 $   & $8s^2 8p^1$ \\
\vspace{9pt} 
122 &                       & $8s_{1/2}^2 8p_{1/2}^1 7d_{3/2}^1             $ &  $2   $   &$\!\begin{aligned}[t]                                                    
                                                                                               &0.85 \cdot 8s_{1/2}^2 8p_{1/2}^1 7d_{3/2}^1  + \\                         
                                                                                               &0.09 \cdot 8s_{1/2}^1 8p_{1/2}^1 7d_{3/2}^2                               
                                                                                               \end{aligned} $  &   $2   $                         &$\!\begin{aligned}[t]
                                                                                                                                                                          &0.84 \cdot 8s_{1/2}^2 8p_{1/2}^1 7d_{3/2}^1  +  \\
                                                                                                                                                                          &0.07 \cdot 8s_{1/2}^1 8p_{1/2}^1 7d_{3/2}^2
                                                                                                                                                                          \end{aligned} $  &   2                 & $7d^1 8p^1$  \\                                                                                                                                                                    
123 &                       & $8s_{1/2}^2 8p_{1/2}^1 7d_{3/2}^1 6f_{5/2}^1$   &  $9/2 $   &$\!\begin{aligned}[t]                                                    
                                                                                               &0.83 \cdot 8s_{1/2}^2 8p_{1/2}^1 7d_{3/2}^1 6f_{5/2}^1 + \\               
                                                                                               &0.14 \cdot 8s_{1/2}^1 8p_{1/2}^1 7d_{3/2}^2 6f_{5/2}^1                    
                                                                                               \end{aligned} $  &   $9/2 $                        &$\!\begin{aligned}[t]
                                                                                                                                                                         &0.82 \cdot 8s_{1/2}^2 8p_{1/2}^1 7d_{3/2}^1 6f_{5/2}^1 + \\
                                                                                                                                                                         &0.09 \cdot 8s_{1/2}^1 8p_{1/2}^1 7d_{3/2}^2 6f_{5/2}^1 
                                                                                                                                                                         \end{aligned} $  &   $9/2 $                   & $6f^2 8p^1$ \\

\midrule                                                                                                                                                            
\vspace{9pt} 
                                                           
$\lvert \underline{\textbf{124}}$ &     +$8s^2$           & $8p_{1/2}^1 7d_{3/2}^1 6f_{5/2}^2             $ &  $6   $   &$\!\begin{aligned}[t]                                                    
                                                                                                &0.83 \cdot 8p_{1/2}^1 6f_{5/2}^3 + \\                              
                                                                                                &0.10 \cdot 8p_{1/2}^1 6f_{5/2}^2 6f_{7/2}^1                        
                                                                                                \end{aligned} $  &   $5   $                       &$\!\begin{aligned}[t]
                                                                                                                                                                          &0.85 \cdot 8p_{1/2}^1 6f_{5/2}^3 + \\
                                                                                                                                                                          &0.07 \cdot 8p_{1/2}^1 6f_{5/2}^2 6f_{7/2}^1
                                                                                                                                                                          \end{aligned} $  &   $5   $                 & $6f^2 8p^2$   \\
\vspace{9pt} 
\textbf{125} &                       & $8p_{1/2}^1 7d_{3/2}^1 6f_{5/2}^2 5g_{7/2}^1  $ &  $17/2$   &$0.96 \cdot 8p_{1/2}^1 7d_{3/2}^1 6f_{5/2}^2 5g_{7/2}^1   $  &   $17/2$  &$0.94 \cdot 8p_{1/2}^1 7d_{3/2}^1 6f_{5/2}^2 5g_{7/2}^1   $  &   $17/2$  &  $\!\begin{aligned}[t]                            
                                                                                                                                                                                                                                                        &0.81 \cdot 5g^1 6f^2 8p^2 + \\      
                                                                                                                                                                                                                                                        &0.17 \cdot 5g^1 6f^1 7d^2 8p^1 + \\
                                                                                                                                                                                                                                                        &0.02 \cdot 6f^3 7d^1 8p^1
                                                                                                                                                                                                                                                        \end{aligned} $ \\
\vspace{9pt} 
\textbf{126} &                       & $8p_{1/2}^1 7d_{3/2}^1 6f_{5/2}^2 5g_{7/2}^2  $ &  $10  $   &$0.93 \cdot 8p_{1/2}^1 7d_{3/2}^1 6f_{5/2}^2 5g_{7/2}^2   $  &   $10  $  &$0.92 \cdot 8p_{1/2}^1 7d_{3/2}^1 6f_{5/2}^2 5g_{7/2}^2   $  &   $10  $  &  $\!\begin{aligned}[t]                       
                                                                                                                                                                                                                                                        &0.998 \cdot 5g^2 6f^3 8p^1 + \\      
                                                                                                                                                                                                                                                        &0.002 \cdot 5g^2 6f^2 8p^2
                                                                                                                                                                                                                                                        \end{aligned} $ \\
\vspace{9pt} 
\textbf{127} &                       & $8p_{1/2}^1 7d_{3/2}^1 6f_{5/2}^2 5g_{7/2}^3  $ &  $27/2$   &$0.95 \cdot 8p_{1/2}^1 7d_{3/2}^1 6f_{5/2}^2 5g_{7/2}^3   $  &   $27/2$   &   $0.94 \cdot 8p_{1/2}^1 7d_{3/2}^1 6f_{5/2}^2 5g_{7/2}^3   $  &   $27/2$   &          $\!\begin{aligned}[t]                   
                                                                                                                                                                                         &0.88 \cdot 5g^3 6f^2 8p^2 + \\ 
                                                                                                                                                                                         &0.12 \cdot 5g^3 6f^1 7d^2 8p^1
                                                                                                                                                                                         \end{aligned} $ \\
\vspace{9pt} 
\textbf{128} &                       & $8p_{1/2}^1 7d_{3/2}^1 6f_{5/2}^2 5g_{7/2}^4  $ &  $14  $   &$\!\begin{aligned}[t] 
                                                                                               &0.80 \cdot 8p_{1/2}^1 7d_{3/2}^1 6f_{5/2}^2 5g_{7/2}^4 +\\
                                                                                               &0.14 \cdot 8p_{1/2}^1 7d_{3/2}^1 6f_{5/2}^2 5g_{7/2}^3 5g_{9/2}^1
                                                                                               \end{aligned}$ &   $14  $                          &$\!\begin{aligned}[t]  
                                                                                                                                                                          &0.81 \cdot 8p_{1/2}^1 7d_{3/2}^1 6f_{5/2}^2 5g_{7/2}^4 +\\
                                                                                                                                                                          &0.13 \cdot 8p_{1/2}^1 7d_{3/2}^1 6f_{5/2}^2 5g_{7/2}^3 5g_{9/2}^1
                                                                                                                                                                          \end{aligned}$ &   $14  $                                 &    $\!\begin{aligned}[t]                                                                                                                                                                                      
                                                                                                                                                                                                                                                               &0.88 \cdot 5g^4 6f^2 8p^2 + \\
                                                                                                                                                                                                                                                               &0.12 \cdot 5g^4 6f^1 7d^2 8p^1
                                                                                                                                                                                                                                                                \end{aligned} $ \\
\vspace{9pt}
\textbf{129} &                       & $8p_{1/2}^1 7d_{3/2}^1 6f_{5/2}^3 5g_{7/2}^4  $ &  $29/2$   &$\!\begin{aligned}[t]
                                                                                               &0.93 \cdot 8p_{1/2}^1 7d_{3/2}^1 6f_{5/2}^3 5g_{7/2}^4  +\\
                                                                                               &0.06 \cdot 8p_{1/2}^1 7d_{3/2}^1 6f_{5/2}^2 5g_{7/2}^4 6f_{7/2}^1
                                                                                               \end{aligned}$  &   $29/2$                          &$0.94 \cdot 8p_{1/2}^1 7d_{3/2}^1 6f_{5/2}^3 5g_{7/2}^4  $  &   $29/2$  & $5g^4 6f^3 7d^1 8p^1$ \\
\vspace{9pt} 
                                                           
$\overline{\textbf{130}} \rvert$ &                       & $8p_{1/2}^1 7d_{3/2}^1 6f_{5/2}^3 5g_{7/2}^5  $ &  $14  $   &$\!\begin{aligned}[t]
                                                                                               &0.92 \cdot 8p_{1/2}^1 7d_{3/2}^1 6f_{5/2}^3 5g_{7/2}^5  + \\
                                                                                               &0.06 \cdot 8p_{1/2}^1 7d_{3/2}^1 6f_{5/2}^2 5g_{7/2}^5 6f_{7/2}^1
                                                                                               \end{aligned}$  &   $14$                             &$\!\begin{aligned}[t] 
                                                                                                                                                                          &0.85 \cdot 8p_{1/2}^2 6f_{5/2}^3 5g_{7/2}^5  + \\
                                                                                                                                                                          &0.06 \cdot 7d_{3/2}^2 6f_{5/2}^3 5g_{7/2}^5
                                                                                                                                                                          \end{aligned}$  &   $12$               &  $5g^5 6f^3 7d^1 8p^1$ \\
\vspace{9pt} 
                                                           
$\lvert \underline{\textbf{131}}$ &                       & $8p_{1/2}^1 7d_{3/2}^1 6f_{5/2}^3 5g_{7/2}^6  $ &  $25/2$   &$\!\begin{aligned}[t]
                                                                                               &0.82 \cdot 8p_{1/2}^2 6f_{5/2}^3 5g_{7/2}^6  + \\
                                                                                               &0.06 \cdot 7d_{3/2}^2 6f_{5/2}^3 5g_{7/2}^6  + \\
                                                                                               &0.06 \cdot 8p_{1/2}^1 7d_{3/2}^2 6f_{5/2}^2 5g_{7/2}^6  +\\
                                                                                               &0.05 \cdot 8p_{1/2}^2 6f_{5/2}^2 5g_{7/2}^6 6f_{7/2}^1
                                                                                               \end{aligned}$  &   $21/2$                         &$\!\begin{aligned}[t]
                                                                                                                                                                          &0.85 \cdot 8p_{1/2}^2 6f_{5/2}^3 5g_{7/2}^6  + \\
                                                                                                                                                                          &0.05 \cdot 7d_{3/2}^2 6f_{5/2}^3 5g_{7/2}^6 
                                                                                                                                                                          \end{aligned}$  &   $21/2$              &  $\!\begin{aligned}[t]                                                                                                      
                                                                                                                                                                                                                                            &0.86 \cdot 5g^6 6f^3 8p^2 + \\
                                                                                                                                                                                                                                            &0.14 \cdot 5g^6 6f^2 7d^2 8p^1
                                                                                                                                                                                                                                             \end{aligned} $ \\
\vspace{9pt} 
                                                           
$\lvert \underline{\textbf{132}}$ &                       & $8p_{1/2}^1 7d_{3/2}^1 6f_{5/2}^3 5g_{7/2}^7  $ &  $10  $   &$\!\begin{aligned}[t]
                                                                                               &0.84 \cdot 8p_{1/2}^2 6f_{5/2}^3 5g_{7/2}^7  + \\
                                                                                               &0.08 \cdot 8p_{1/2}^1 7d_{3/2}^2 6f_{5/2}^2 5g_{7/2}^7  + \\
                                                                                               &0.06 \cdot 7d_{3/2}^2 6f_{5/2}^3 5g_{7/2}^7
                                                                                               \end{aligned}$  &   $6   $                          &$\!\begin{aligned}[t]
                                                                                                                                                                          &0.87 \cdot 8p_{1/2}^2 6f_{5/2}^3 5g_{7/2}^7  + \\
                                                                                                                                                                          &0.05 \cdot 7d_{3/2}^2 6f_{5/2}^3 5g_{7/2}^7
                                                                                                                                                                          \end{aligned}$  &   $6   $                 &  $5g^7 6f^3 8p^2$ \\
                                                           
$\lvert \underline{\textbf{133}}$ &                       & $8p_{1/2}^1 7d_{3/2}^1 6f_{5/2}^3 5g_{7/2}^8  $ &  $13/2$   &$\!\begin{aligned}[t]
                                                                                               &0.83 \cdot 8p_{1/2}^2 6f_{5/2}^3 5g_{7/2}^8  + \\
                                                                                               &0.09 \cdot 8p_{1/2}^1 7d_{3/2}^2 6f_{5/2}^2 5g_{7/2}^8
                                                                                               \end{aligned}$  &   $9/2 $                           &$0.87 \cdot 8p_{1/2}^2 6f_{5/2}^3 5g_{7/2}^8$     & $9/2$        &  $5g^8 6f^3 8p^2$ \\
\midrule
\vspace{9pt} 
                                                           
$\lvert \underline{\textbf{134}}$ &      +$5g_{7/2}^8$    & $8p_{1/2}^1 7d_{3/2}^1 6f_{5/2}^4             $ &  $6   $   &$\!\begin{aligned}[t]
                                                                                               &0.82 \cdot 8p_{1/2}^2 6f_{5/2}^4  + \\
                                                                                               &0.07 \cdot 8p_{1/2}^1 7d_{3/2}^2 6f_{5/2}^3  + \\
                                                                                               &0.06 \cdot 7d_{3/2}^2 6f_{5/2}^4
                                                                                               \end{aligned}$  &   $4   $                          &$\!\begin{aligned}[t]
                                                                                                                                                                           &0.84 \cdot 8p_{1/2}^2 6f_{5/2}^4  + \\
                                                                                                                                                                           &0.05 \cdot 8p_{1/2}^1 7d_{3/2}^2 6f_{5/2}^3  + \\
                                                                                                                                                                           &0.05 \cdot 7d_{3/2}^2 6f_{5/2}^4
                                                                                                                                                                           \end{aligned}$  &   $4   $                &  $5g^8 6f^4 8p^2$ \\                                                                                                                                                                                                                                                    
\vspace{9pt}                                                                                                                                                                            
                                                           
$\lvert \underline{\textbf{135}}$ &                       & $8p_{1/2}^2 7d_{3/2}^1 6f_{5/2}^3 5g_{9/2}^1  $ &  $13/2$   &$\!\begin{aligned}[t]
                                                                                               &0.82 \cdot 8p_{1/2}^2 6f_{5/2}^4 5g_{9/2}^1  + \\
                                                                                               &0.08 \cdot 8p_{1/2}^1 7d_{3/2}^2 6f_{5/2}^3 5g_{9/2}^1  + \\
                                                                                               &0.05 \cdot 7d_{3/2}^2 6f_{5/2}^4 5g_{9/2}^1
                                                                                               \end{aligned}$  &   $5/2 $                          &$\!\begin{aligned}[t]
                                                                                                                                                                          &0.85 \cdot 8p_{1/2}^2 6f_{5/2}^4 5g_{9/2}^1  + \\
                                                                                                                                                                          &0.06 \cdot 8p_{1/2}^1 7d_{3/2}^2 6f_{5/2}^3 5g_{9/2}^1  
                                                                                                                                                                          \end{aligned}$  &   $5/2 $                 &  $5g^9 6f^4 8p^2$ \\                                                                                                                                                                                            
\vspace{9pt} 
$\underline{\textbf{136}}$ &                              & $8p_{1/2}^2 7d_{3/2}^1 6f_{5/2}^3 5g_{9/2}^2             $ &  $3   $   &$\!\begin{aligned}[t]
                                                                                               &0.91 \cdot 8p_{1/2}^2 7d_{3/2}^1 6f_{5/2}^3 5g_{9/2}^2 +\\
                                                                                               &0.05 \cdot 7d_{3/2}^3 6f_{5/2}^3 5g_{9/2}^2 \\
                                                                                               \end{aligned}$&   $4$  &       $0.89 \cdot 8p_{1/2}^2 7d_{3/2}^1 6f_{5/2}^3 5g_{9/2}^2       $  & $4$  &  $5g^{10} 6f^4 8p^2$ \\
\vspace{9pt} 
$\lvert \overline{\underline{137}} \rvert$ &                       & $8p_{1/2}^2 7d_{3/2}^1 6f_{5/2}^3 5g_{9/2}^3             $ &  $19/2$   &$\!\begin{aligned}[t]
                                                                                               &0.80 \cdot  8p_{1/2}^2 6f_{5/2}^4 5g_{9/2}^3  + \\
                                                                                               &0.11 \cdot  8p_{1/2}^1 7d_{3/2}^2 6f_{5/2}^3 5g_{9/2}^3   \\
                                                                                               \end{aligned}$  &   $13/2 $       & $0.89 \cdot 8p_{1/2}^2 7d_{3/2}^1 6f_{5/2}^3 5g_{9/2}^3       $  & $17/2$   &  $5g^{11} 6f^4 8p^2$ \\
\midrule
\vspace{9pt} 
$\underline{138}$ &        +$8p_{1/2}^2$       & $7d_{3/2}^1 6f_{5/2}^3 5g_{9/2}^4             $ &  $6   $   & $0.91 \cdot 7d_{3/2}^1 6f_{5/2}^3 5g_{9/2}^4              $  &   $7   $  & $0.89 \cdot 7d_{3/2}^1 6f_{5/2}^3 5g_{9/2}^4       $  & $7$                                                                 &  $5g^{12} 6f^3 7d^1 8p^2$ \\
\vspace{9pt} 
139 &                       & $7d_{3/2}^1 6f_{5/2}^3 5g_{9/2}^5             $ &  $13/2$   & $0.92 \cdot 7d_{3/2}^1 6f_{5/2}^3 5g_{9/2}^5              $  &   $13/2$  & $0.91 \cdot 7d_{3/2}^1 6f_{5/2}^3 5g_{9/2}^5       $  & $13/2$                                                                 &  $5g^{13} 6f^3 7d^1 8p^2$ \\
\vspace{9pt} 
$\lvert \textbf{140}$ &                       & $7d_{3/2}^2 6f_{5/2}^2 5g_{9/2}^6             $ &  $6   $   &$\!\begin{aligned}[t]
                                                                                               &0.90 \cdot 7d_{3/2}^1 6f_{5/2}^3 5g_{9/2}^6 + \\
                                                                                               &0.05 \cdot 7d_{3/2}^1 6f_{5/2}^2 5g_{9/2}^6 6f_{7/2}^1
                                                                                               \end{aligned}$  &   $6   $           &$0.90 \cdot 7d_{3/2}^1 6f_{5/2}^3 5g_{9/2}^6$ &   $6$                                                                &  $5g^{14} 6f^3 7d^1 8p^2$ \\
\vspace{9pt} 
141 &                       & $7d_{3/2}^2 6f_{5/2}^2 5g_{9/2}^7             $ &  $9/2 $   & $0.93 \cdot 7d_{3/2}^2 6f_{5/2}^2 5g_{9/2}^7              $  &   $9/2 $  &    $ 0.91\cdot 7d_{3/2}^2 6f_{5/2}^2 5g_{9/2}^7  $ &  $9/2 $                                                           &  $5g^{15} 6f^2 7d^2 8p^2$ \\
\vspace{9pt} 
142 &                       & $7d_{3/2}^2 6f_{5/2}^2 5g_{9/2}^8             $ &  $2   $   & $0.91 \cdot 7d_{3/2}^2 6f_{5/2}^2 5g_{9/2}^8              $  &   $2   $  & $0.89 \cdot 7d_{3/2}^2 6f_{5/2}^2 5g_{9/2}^8              $  &   $2   $                                                               &  $5g^{16} 6f^2 7d^2 8p^2$ \\
\vspace{9pt} 
$\overline{\underline{\textbf{143}}}$ &                       & $7d_{3/2}^2 6f_{5/2}^3 5g_{9/2}^8             $ &  $5/2 $   & $0.93 \cdot 7d_{3/2}^2 6f_{5/2}^3 5g_{9/2}^8              $  &   $3/2 $  & $0.92 \cdot 7d_{3/2}^2 6f_{5/2}^3 5g_{9/2}^8       $  & $1/2$                                      &  $5g^{17} 6f^2 7d^2 8p^2$ \\
\vspace{9pt} 
\textbf{144} &                       & $7d_{3/2}^2 6f_{5/2}^3 5g_{9/2}^9             $ &  $7   $   & $0.96 \cdot 7d_{3/2}^2 6f_{5/2}^3 5g_{9/2}^9              $  &   $7   $  & $0.94 \cdot 7d_{3/2}^2 6f_{5/2}^3 5g_{9/2}^9              $  &   $7   $                                                                &  $\!\begin{aligned}[t]
                                                                                                                                                                                                                                                 &0.95 \cdot 5g^{17} 6f^2 7d^3 8p^2 + \\
                                                                                                                                                                                                                                                 &0.05 \cdot 5g^{17} 6f^4 7d^1 8p^2
                                                                                                                                                                                                                                                  \end{aligned} $ \\
145 &                       & $7d_{3/2}^2 6f_{5/2}^3 5g_{9/2}^{10}          $ &  $13/2$   & $0.96 \cdot 7d_{3/2}^2 6f_{5/2}^3 5g_{9/2}^{10}           $  &   $13/2$  & $0.93 \cdot 7d_{3/2}^2 6f_{5/2}^3 5g_{9/2}^{10}           $  &   $13/2$                    &  $5g^{18} 6f^3 7d^2 8p^2$ \\
\midrule
\vspace{9pt} 
146 &      +$5g_{9/2}^{10}$ & $7d_{3/2}^2 6f_{5/2}^4                        $ &  $6   $   & $0.95 \cdot 7d_{3/2}^2 6f_{5/2}^4                         $  &   $6   $  & $0.91 \cdot 7d_{3/2}^2 6f_{5/2}^4                         $  &   $6   $                       &  $6f^4 7d^2 8p^2$ \\
\vspace{9pt}                                                                                                                                                                          
$\underline{147}$ &                       & $7d_{3/2}^2 6f_{5/2}^5                        $ &  $7/2 $   &$\!\begin{aligned}[t]
                                                                                            &0.89 \cdot 7d_{3/2}^2 6f_{5/2}^5  + \\
                                                                                            &0.07 \cdot 7d_{3/2}^2 6f_{5/2}^4 6f_{7/2}^1
                                                                                            \end{aligned}$  &   $9/2 $                            &$\!\begin{aligned}[t]
                                                                                                                                                                       &0.88 \cdot 7d_{3/2}^2 6f_{5/2}^5  + \\
                                                                                                                                                                       &0.05 \cdot 7d_{3/2}^2 6f_{5/2}^4 6f_{7/2}^1
                                                                                                                                                                       \end{aligned}$  &   $9/2 $                   &  $6f^5 7d^2 8p^2$ \\                                                                                                                                                                                                                                                                                                                  
\vspace{9pt}                                                                                            
148 &                       & $7d_{3/2}^2 6f_{5/2}^6                        $ &  $2   $   & $0.94 \cdot 7d_{3/2}^2 6f_{5/2}^6                         $  &   $2   $  & $0.93 \cdot 7d_{3/2}^2 6f_{5/2}^6                         $  &   $2   $   &  $6f^6 7d^2 8p^2$ \\  
\vspace{9pt}
149 &                       & $7d_{3/2}^3 6f_{5/2}^6                        $ &  $3/2 $   & $0.96 \cdot 7d_{3/2}^3  6f_{5/2}^6                                    $  & $3/2 $  & $0.93 \cdot 7d_{3/2}^3  6f_{5/2}^6              $  &   $3/2 $  &  $6f^6 7d^3 8p^2$ \\  
\vspace{9pt}
\textbf{150} &                       & $7d_{3/2}^3 6f_{5/2}^6 6f_{7/2}^1             $ &  $2   $   &$\!\begin{aligned}[t]
                                                                                                   &0.85 \cdot 7d_{3/2}^3 6f_{5/2}^6 6f_{7/2}^1  + \\
                                                                                                   &0.11 \cdot 7d_{3/2}^3 6f_{5/2}^5 6f_{7/2}^2
                                                                                                   \end{aligned}$  &   2                           &$\!\begin{aligned}[t]
                                                                                                                                                                              &0.89 \cdot 7d_{3/2}^3 6f_{5/2}^6 6f_{7/2}^1  + \\
                                                                                                                                                                              &0.06 \cdot 7d_{3/2}^3 6f_{5/2}^5 6f_{7/2}^2
                                                                                                                                                                              \end{aligned}$  &   2           &  $6f^7 7d^3 8p^2$ \\  
151 &                       & $7d_{3/2}^3 6f_{5/2}^6 6f_{7/2}^2             $ &  $9/2 $   &$\!\begin{aligned}[t]
                                                                                               &0.89 \cdot 7d_{3/2}^3 6f_{5/2}^6 6f_{7/2}^2  + \\
                                                                                               &0.09 \cdot 7d_{3/2}^3 6f_{5/2}^5 6f_{7/2}^3
                                                                                               \end{aligned}$  &   $9/2 $                       &$\!\begin{aligned}[t]
                                                                                                                                                                         &0.89 \cdot 7d_{3/2}^3 6f_{5/2}^6 6f_{7/2}^2  + \\
                                                                                                                                                                         &0.06 \cdot 7d_{3/2}^3 6f_{5/2}^5 6f_{7/2}^3
                                                                                                                                                                         \end{aligned}$  &   $9/2 $               &  $6f^8 7d^3 8p^2$ \\                                                                                                
\midrule    
\vspace{9pt}
\textbf{152} &     +$6f_{5/2}^{6}$   & $7d_{3/2}^2 6f_{7/2}^4                        $ &  $6   $   & $0.95 \cdot 7d_{3/2}^2 6f_{7/2}^4                         $  &   $6   $     & $0.92 \cdot 7d_{3/2}^2 6f_{7/2}^4                         $  &   $6   $   &  $6f^9 7d^3 8p^2$ \\   
\vspace{9pt}
153 &                       & $7d_{3/2}^2 6f_{7/2}^5                        $ &  $11/2$   & $0.96 \cdot 7d_{3/2}^2 6f_{7/2}^5                         $  &   $11/2$     & $0.93 \cdot 7d_{3/2}^2 6f_{7/2}^5                         $  &   $11/2$   &  $6f^{10} 7d^3 8p^2$ \\   
\vspace{9pt}
154 &                       & $7d_{3/2}^2 6f_{7/2}^6                        $ &  $6   $   & $0.98 \cdot 7d_{3/2}^2 6f_{7/2}^6                         $  &   $6   $     & $0.95 \cdot 7d_{3/2}^2 6f_{7/2}^6                         $  &   $6   $   &  $6f^{11} 7d^3 8p^2$ \\   
\vspace{9pt}
$\lvert \underline{\textbf{155}}$ &                       & $7d_{3/2}^2 6f_{7/2}^7                        $ &  $7/2 $   & $0.99 \cdot 9s_{1/2}^1 6f_{7/2}^8                          $  &   $1/2 $    & $0.94 \cdot 9s_{1/2}^1 6f_{7/2}^8                          $  &   $1/2 $  &  $6f^{12} 7d^3 8p^2$ \\   
\vspace{9pt}
156 &                       & $7d_{3/2}^2 6f_{7/2}^8                        $ &  $2   $   & $0.97 \cdot 7d_{3/2}^2 6f_{7/2}^8                         $  &   $2   $     & $0.97 \cdot 7d_{3/2}^2 6f_{7/2}^8                         $  &   $2   $   &  $6f^{13} 7d^3 8p^2$ \\   
\midrule                                                                                                                                                                       
\vspace{12pt}
157 &      +$6f_{7/2}^8$    & $7d_{3/2}^3                                   $ &  $3/2 $   & $0.96 \cdot 7d_{3/2}^3                                    $  &   $3/2 $     & $0.96 \cdot 7d_{3/2}^3                                    $  &   $3/2 $   &  $6f^{14} 7d^3 8p^2$ \\ 
\vspace{9pt}
158 &                       & $7d_{3/2}^4                                   $ &  $0   $   & $0.98 \cdot 7d_{3/2}^4                                    $  &   $0   $     & $0.96 \cdot 7d_{3/2}^4                                    $  &   $0   $   &  $6f^{14} 7d^4 8p^2$ \\   
\midrule    
\vspace{9pt}
159 &      +$7d_{3/2}^4$    & $9s_{1/2}^1                                   $ &  $1/2 $   & $0.98 \cdot 9s_{1/2}^1                                    $  &   $1/2 $     & $0.96 \cdot 9s_{1/2}^1                                    $  &   $1/2 $   &  $6f^{14} 7d^4 8p^2 9s^1$ \\
\vspace{9pt}
160 &                       & $7d_{5/2}^1 9s_{1/2}^1                        $ &  $3   $   & $0.96 \cdot 7d_{5/2}^1 9s_{1/2}^1                         $  &   $3   $     & $0.95 \cdot 7d_{5/2}^1 9s_{1/2}^1                         $  &   $3   $   &  $6f^{14} 7d^5 8p^2 9s^1$ \\
\vspace{9pt}
161 &                       & $7d_{5/2}^2 9s_{1/2}^1                        $ &  $9/2 $   & $0.97 \cdot 7d_{5/2}^2 9s_{1/2}^1                         $  &   $9/2 $     & $0.92 \cdot 7d_{5/2}^2 9s_{1/2}^1                         $  &   $9/2 $   &  $6f^{14} 7d^6 8p^2 9s^1$ \\
\vspace{9pt}
\textbf{162} &                       & $7d_{5/2}^3 9s_{1/2}^1                        $ &  $5   $   & $0.98 \cdot 7d_{5/2}^3 9s_{1/2}^1                         $  &   $5   $     & $0.96 \cdot 7d_{5/2}^3 9s_{1/2}^1                         $  &   $5   $   &  $6f^{14} 7d^7 8p^2 9s^1$ \\
\vspace{9pt}
163 &                       & $7d_{5/2}^5                                   $ &  $5/2 $   & $0.96 \cdot 7d_{5/2}^5                                    $  &   $5/2 $     & $0.95 \cdot 7d_{5/2}^5                                    $  &   $5/2 $   &  $6f^{14} 7d^8 8p^2 9s^1$ \\
\vspace{9pt}
164 &                       & $7d_{5/2}^6                                   $ &  $0   $   & $0.98 \cdot 7d_{5/2}^6                                    $  &   $0   $     & $0.96 \cdot 7d_{5/2}^6                                    $  &   $0   $   &  $6f^{14} 7d^{10} 8p^2$  \\
\midrule                                                                                                                                                                       
\vspace{9pt}
165 &      +$7d_{5/2}^6$    & $9s_{1/2}^1                                   $ &  $1/2 $   & $0.98 \cdot 9s_{1/2}^1                                    $  &   $1/2 $     & $0.96 \cdot 9s_{1/2}^1                                    $  &   $1/2 $   &         \\
\vspace{9pt}
166 &                       & $9s_{1/2}^2                                   $ &  $0   $   & $0.84 \cdot 9s_{1/2}^2  + 0.10 \cdot 8p_{3/2}^2$ &   $0   $  & $0.90 \cdot 9s_{1/2}^2$ &   $0   $                              &        \\
\vspace{9pt}
$\lvert \textbf{167}$ &                       & $9s_{1/2}^1 8p_{3/2}^1  9p_{1/2}^1                       $ &  $3/2 $   &$0.88 \cdot 9s_{1/2}^2 8p_{3/2}^1 + 0.06 \cdot 8p_{3/2}^3$ &   $3/2 $               & $0.91 \cdot 9s_{1/2}^2 8p_{3/2}^1$ & $3/2$                        &         \\
\vspace{9pt}
$\lvert \overline{\underline{168}} \rvert$ &                       & $9s_{1/2}^2 8p_{3/2}^1 9p_{1/2}^1             $ &  $1   $   &$\!\begin{aligned}[t]
                                                                                               &0.55 \cdot 9s_{1/2}^1 8p_{3/2}^2 9p_{1/2}^1  + \\
                                                                                               &0.24 \cdot 9s_{1/2}^1 8p_{3/2}^1 9p_{1/2}^2  + \\
                                                                                               &0.20 \cdot 9s_{1/2}^1 8p_{3/2}^3  
                                                                                               \end{aligned}$  &   $2$                                         & $0.94 \cdot 9s_{1/2}^2 8p_{3/2}^1 9p_{1/2}^1$ & $1$                & \\
\vspace{9pt}
$\textbf{169}$ &                       & $9s_{1/2}^2 8p_{3/2}^2 9p_{1/2}^1             $ &  $3/2 $   &$\!\begin{aligned}[t]
                                                                                               &0.76 \cdot 9s_{1/2}^2 8p_{3/2}^2 9p_{1/2}^1  + \\
                                                                                               &0.16 \cdot 9s_{1/2}^2 8p_{3/2}^1 9p_{1/2}^2  + \\
                                                                                               &0.06 \cdot 9s_{1/2}^2 8p_{3/2}^3
                                                                                               \end{aligned}$  &   $3/2$    \                      &$\!\begin{aligned}[t]
                                                                                                                                                                          &0.83 \cdot 9s_{1/2}^2 8p_{3/2}^2 9p_{1/2}^1  + \\
                                                                                                                                                                          &0.07 \cdot 9s_{1/2}^2 8p_{3/2}^1 9p_{1/2}^2  
                                                                                                                                                                          \end{aligned}$  &   $3/2$  & \\                                                                                                                                                                                                                                                                                                                                                                                                                                                                                 
170 &                       & $9s_{1/2}^2 8p_{3/2}^2 9p_{1/2}^2                        $ &  $2   $   &
                                                                                                          $0.96 \cdot 9s_{1/2}^2 8p_{3/2}^2 9p_{1/2}^2$ 
                                                                                                           &   $2$           
                                                                                                                                                                         &$0.93 \cdot 9s_{1/2}^2 8p_{3/2}^2 9p_{1/2}^2$  &   $2   $  & \\

\bottomrule
\end{longtable}
\label{tab:6}
\twocolumngrid

\end{document}